\documentclass[twocolumn,superscriptaddress,aps,prb,showkeys]{revtex4-2}
\usepackage{color}
\usepackage{amssymb}
\usepackage[normalem]{ulem}

\usepackage{amsmath}
\usepackage{graphicx}	

\usepackage{todonotes}

\usepackage[colorlinks=true]{hyperref}	

\usepackage{bibunits}

\begin{document}

	\title{Distributed Current Injection into a One-Dimensional Ballistic Edge Channel}
	
	\author{Kristof Moors}
	\thanks{Present Address: Imec, Kapeldreef 75, 3001 Leuven, Belgium}
	\email[]{kristof.moors@imec.be}
	\affiliation{Peter Gr\"unberg Institute (PGI-9), Forschungszentrum J\"ulich, 52425 J\"ulich, Germany}
	\affiliation{Jülich Aachen Research Alliance (JARA), Fundamentals of Future Information Technology, 52425 Jülich, Germany}
	\author{Christian Wagner}
	\affiliation{Peter Gr\"unberg Institute (PGI-3), Forschungszentrum J\"ulich, 52425 J\"ulich, Germany}
	\author{Helmut Soltner}
	\affiliation{Institute of Technology and Engineering (ITE), Forschungszentrum J\"ulich, 52425 J\"ulich, Germany}
	\author{Felix~L\"upke}
 	\affiliation{Jülich Aachen Research Alliance (JARA), Fundamentals of Future Information Technology, 52425 Jülich, Germany}
	\affiliation{Peter Gr\"unberg Institute (PGI-3), Forschungszentrum J\"ulich, 52425 J\"ulich, Germany}
	\affiliation{Institute of Physics II, Universit\"at zu K\"oln, Z\"ulpicher Straße 77, 50937 K\"oln, Germany}
	\author{F.~Stefan Tautz}\email[]{s.tautz@fz-juelich.de}
	\affiliation{Peter Gr\"unberg Institute (PGI-3), Forschungszentrum J\"ulich, 52425 J\"ulich, Germany}
	\affiliation{Jülich Aachen Research Alliance (JARA), Fundamentals of Future Information Technology, 52425 Jülich, Germany}
	\affiliation{Experimental Physics IV A, RWTH Aachen University, Otto-Blumenthal-Straße, 52074 Aachen, Germany}
	\author{Bert Voigtl\"ander}
	\affiliation{Peter Gr\"unberg Institute (PGI-3), Forschungszentrum J\"ulich, 52425 J\"ulich, Germany}
	\affiliation{Jülich Aachen Research Alliance (JARA), Fundamentals of Future Information Technology, 52425 Jülich, Germany}
	\affiliation{Experimental Physics IV A, RWTH Aachen University, Otto-Blumenthal-Straße, 52074 Aachen, Germany}
	
	\date{\today}
	
	\begin{abstract}
    We generalize Landauer's theory of ballistic transport in a one-dimensional (1D) conductor to situations where charge carrier injection and extraction are not any more confined to electrodes at either end of the channel, but may occur along its whole length. This type of distributed injection is expected to occur from the two-dimensional (2D) bulk of, e.g., a quantum spin (or anomalous) Hall insulator to its topologically protected edge states.
    We apply our conceptual solution to the case of two metal electrodes contacting the 2D bulk, enabling us to derive criteria that discriminate ballistic from resistive edge channels in multi-terminal transport experiments.
    \end{abstract}
	
	
	\maketitle
    
    \begin{bibunit}[]
    
	\begin{figure*}[tb]
		\includegraphics[width=\linewidth]{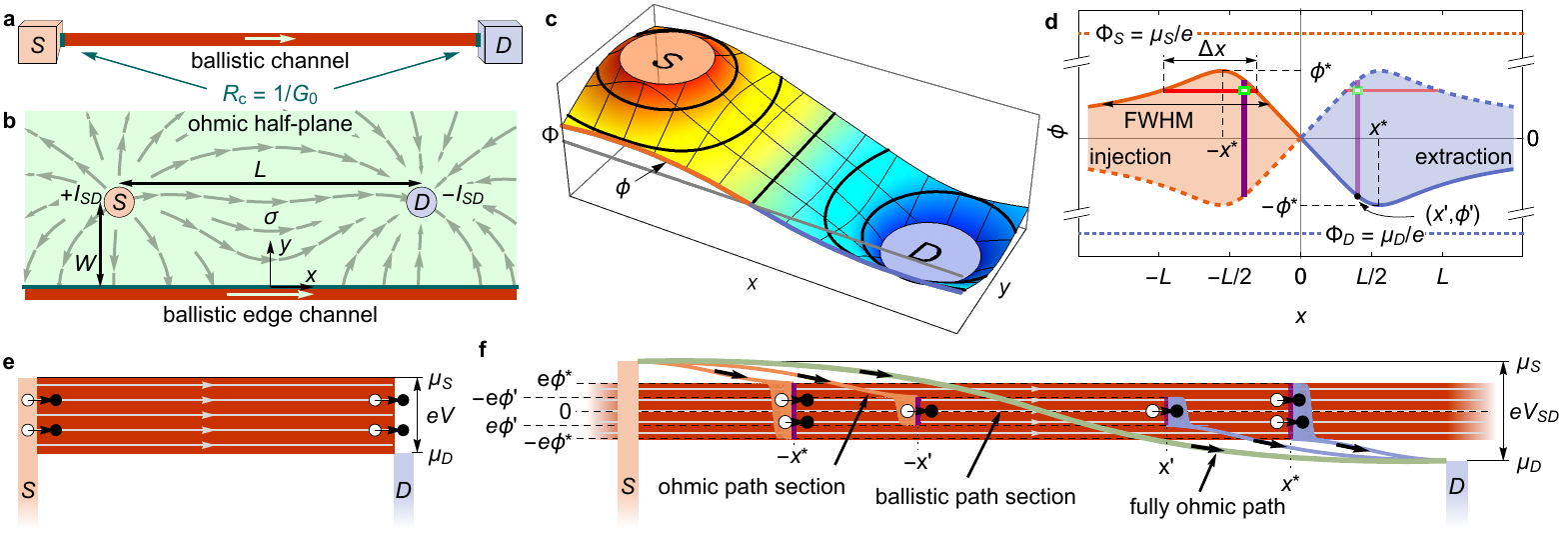}
		\caption{\textbf{1D ballistic edge channel with distributed injection.}
			\textbf{(a)} In the standard Landauer setup, charge carriers are injected and extracted locally at the ends of a 1D ballistic channel. The contact resistance at each contact (indicated by the teal interfaces) is $R_\mathrm{c} = 1/G_0$ per spin.
            \textbf{(b)} In the contact geometry considered here, current enters and exits a 1D ballistic edge channel in a distributed manner via a conducting half-plane with sheet conductivity $\sigma$ and symmetrically positioned source (S) and drain (D) contacts (e.g., STM tips).
			\textbf{(c)} Potential $\Phi(x,y)$ in the 2D half-plane, obtained by solving the Poisson equation with boundary condition $\Phi(x,y=0^+) = \phi(x)$ at the interface with the 1D ballistic channel. The interface potential $\phi(x)$, indicated as solid orange (blue) lines on the source (drain) side, is obtained by numerically solving the filling condition.
			\textbf{(d)} The antisymmetric $\phi(x)$ (see text for details).
            \textbf{(e)} Energy diagram of charge carrier transport in the standard Landauer setup shown in (a).
            \textbf{(f)} Energy diagram for the contact geometry shown in (b).
		}
		\label{fig:1}
	\end{figure*}
	
    Charge transport through a 1D channel without scattering was famously explained by Rolf Landauer decades ago~\cite{Landauer1970,Landauer1981,Landauer1996}. He considered a ballistic conductor between two metal electrodes, with charge carrier injection at one end and extraction at the other (Fig.~\ref{fig:1}a). Importantly, in Landauer's seminal work neither injection nor extraction is allowed along the length of the channel. This situation is realized, e.g., in 2D electron gases in GaAs-AlGaAs heterostructures~\cite{vanWees1988,Picciotto2001} and carbon nanotubes~\cite{Frank1998}, for which Landauer's theory has provided accurate predictions.	
 
    With the recently rising interest in topologically protected edge channels in quantum spin or anomalous Hall systems (semiconductor quantum wells~\cite{klitzing1980new, konig2007quantum, roth2009nonlocal, knez2011evidence, lai2011imaging, Nowack2013, fijalkowski2021quantum}, 2D materials~\cite{novoselov2005two, chang2013experimental, pauly2015subnanometre, reis2017bismuthene, Fei2017,Wu2018, shi2019imaging, allen2019visualization, lippertz2022current, ferguson2023direct, johnsen2023mapping}, and graphene nanoribbons~\cite{Baringhaus2014,Aprojanz2018}), a fundamentally different experimental situation has moved into focus: the ballistic channel exists alongside a 2D half-plane with which it forms an interface along the complete channel length; beyond this extended interface contact, there are no further specific injection contacts. If, as is the case in most experimental realizations of the systems mentioned above, the half-plane has a nonvanishing residual conductivity, any voltage applied within the half-plane will cause a current injection into the ballistic channel that is distributed along the length of the channel. Clearly, this set of circumstances is not covered by Landauer's original considerations~\cite{Landauer1970,Landauer1981,Landauer1996}.

    Here, we generalize Landauer's theory of ballistic transport to the situation of distributed injection into a 1D channel. To this end, we specify an injection (and extraction) current density distribution in the 2D half-plane and solve the transport problem from the fundamental principle of conductance quantization in the 1D channel with perfect transmission (conductance  $G_0/2=e^2/h\approx(25.8\,\mathrm{k\Omega})^{-1}$ per spin, e.g.,~Ref.~\cite{Datta1995}).
    We consider the transport in the contacts (here: the 2D half-plane) as diffusive and thus employ a semiclassical approach based on Poisson's equation~\cite{Leis2022b}, rather than a fully self-consistent quantum mechanical treatment~\cite{Armagnat2019, Flor2022}.

    Beyond formulating a conceptual, geometry-independent framework for distributed injection, we solve a specific class of injection geometries---symmetrically placed point-like injection contacts in the half-plane. This geometry is directly applicable to nanoprobes and multi-tip scanning tunneling microscopy (STM), powerful experimental methods to study 2D transport at the nanoscale~\cite{Voigtlander2018}. Using the explicit solution provided here, multi-terminal potentiometry~\cite{Luepke2015} can overcome the challenge to discriminate topological (ballistic) from trivial (diffusive) edge channels, which is impossible in single-tip STM, because in both cases the vertical tunnelling conductivity can be enhanced by the high local density of states near the edge~\cite{pauly2015subnanometre, reis2017bismuthene, tang2017quantum, lupke2022quantum, lupke2022local,johnsen2023mapping, Yu2024}.
	
    \textit{Contact geometry}.---For current injection into the conducting 2D half-plane with sheet conductivity $\sigma$, we consider small circular source (S, potential $\Phi_{S}=\mu_{S}/e$) and drain (D, potential $\Phi_{D}=\mu_{D}/e$) contacts (Fig.~\ref{fig:1}b). This situation can be realized, for instance, with small lithographic contacts or by contacting two tips of a multi-tip STM to a sample surface~\cite{Voigtlander2018,Leis2021,Leis2022,Leis2022b}.
    Source and drain are positioned at a distance $L$ from each other and a common distance $W$ from the ballistic channel. For the configuration in Fig.~\ref{fig:1}b, the whole region $x < 0$, $y=0$ can be considered as the region of distributed injection from the source contact into the 1D ballistic channel and, correspondingly, the region $x > 0$, $y=0$ as the region of distributed extraction towards the drain contact.
 
    \textit{Distributed injection}.--- In the situation displayed in Fig.~\ref{fig:1}b, a current distribution develops over the half-plane with different types of current paths from source to drain: (i) fully Ohmic paths that do not enter the ballistic channel and (ii) paths that go via the ballistic channel, thereby including both resistive and ballistic sections.
    The current paths can be obtained from a continuously varying potential $\Phi(x, y)$ whose profile is governed by the Poisson equation ($\nabla^2 \Phi(x,y) = -\nabla \cdot \mathbf{j}/\sigma$) with appropriate boundary conditions for the contacts and the interface with the ballistic channel at $y = 0$. We consider, without loss of generality, anti-symmetric boundary conditions at the contacts, giving rise to an anti-symmetric potential along $x$, $\Phi(x, y) = -\Phi(-x, y)$.
    The boundary condition for the potential at the interface to the ballistic channel depends on the latter's properties (more specifically, the filling of its states) which will be discussed below. It results in an interface potential profile $\Phi(x, y=0^+) \equiv \phi(x)$, the generic shape of which, including a single maximum $\phi^\ast$ and $\phi(x) = -\phi(-x)$ (see solid orange and blue lines in Fig.~\ref{fig:1}c-d), follows from symmetry considerations (\ref{sec:appendixA}).
	
    Utilizing the generic $\phi(x)$ in Fig.~\ref{fig:1}d, we discuss charge carrier injection into and extraction out of the ballistic channel with the help of Figs.~\ref{fig:1}e-f. In Landauer's treatment, a source reservoir injects right movers from an energy window $[\mu_{D},\mu_{S}]$ into the ballistic channel (Fig.~\ref{fig:1}e), corresponding to a bias voltage $V = \Phi_{S}-\Phi_{D} = (\mu_{S}-\mu_{D})/e$ (here considering positive charge carriers and $V>0$ for convenience, without loss of generality). For the ballistic channel to exhibit a perfectly quantized conductance, all available right moving states in the channel must be occupied through the injection in this energy window and propagate with perfect transmission between source and drain, without being compensated by states moving in the opposite direction (reflectionless drain contact)~\cite{Datta1995}.	
    In contrast, in the distributed contact geometry the energy window for injection becomes a function of $x$, as shown by the orange shaded region in Fig.~\ref{fig:1}d. On each Ohmic current path section between source contact and ballistic channel, carriers lower their energy by $\mu_{S}-e\phi(x)$ (light orange paths with arrows in Fig.~\ref{fig:1}f for two different values of $x$). This yields $e \phi(x)$ as a local upper limit for the energy of carriers injected at $x$ into the ballistic channel on the source side ($x<0$), shown as solid orange line in Fig.~\ref{fig:1}d. A similar argument applies to the carriers extracted on the drain side ($x>0$) that lower their energy by $e\phi(x) - \mu_{D}$ (light blue paths in Fig.~\ref{fig:1}f) between the ballistic channel and the drain contact. As they can only enter the drain contact with energy $\mu_{D}$ or higher, the lower energy limit to exit the ballistic channel at $x>0$ is $e \phi(x)$ (solid blue line in Fig.~\ref{fig:1}d). Dictated by symmetry, corresponding injection and extraction paths, arriving at the ballistic channel at $x<0$ and departing from it at $-x>0$, respectively, feature the same energy loss in the 2D plane. As a consequence, the energy distribution of injected charge carriers at $x<0$ must be identical to that of the extracted carriers at $-x>0$. This implies that the upper (lower) limit for injection (extraction) at $x<0$ ($x>0$) is also the upper (lower) limit of extraction (injection) at $-x>0$ ($-x<0$), shown as dashed blue (orange) line in Fig.~\ref{fig:1}d (\ref{sec:derivation}). Hence, at all $x$, injection and extraction windows at the ballistic channel are spread symmetrically around zero energy with width $2e|\phi(x)|$ and symmetric around $x=0$ (see orange and blue shaded regions in Fig.~\ref{fig:1}d). 

    When charge carriers enter the ballistic channel, they leave holes behind in the Ohmic 2D plane (indicated by white circles at positions $-x^\ast$ and $-x'$ in Fig.~\ref{fig:1}f). These holes are highly energetic compared to the surrounding Fermi sea and quickly filled by dissipative relaxation processes in the 2D Ohmic plane within a short distance $\lambda_\mathrm{mfp}$, as indicated by orange areas close to the interface in Fig.~\ref{fig:1}f. We consider $\lambda_\mathrm{mfp} \ll |\Phi_{S}|/|\nabla\Phi|, L, W$ such that conventional Ohmic transport with sheet conductivity $\sigma$ is effectively maintained in the half-plane and $\lambda_\text{mfp}$ does not enter our solutions explicitly (\ref{sec:Analysisfillingcondition}). Analogous processes occur when charge carriers exit the ballistic channel. Together, these relaxation processes add to the overall dissipation for a path from the source to the drain electrode via the ballistic channel, such that the total dissipation is equal to $\mu_{S}-\mu_{D}$ for all current paths ($2e|\phi(x')|$ occurring within $\lambda_\mathrm{mfp}$ of the interface to the channel, and $\mu_S-\mu_D - 2e|\phi(x')|$ spread over the entire path through the 2D half-plane, for a current path with entry point $-x'$ and exit point $x'$). 

\begin{figure}[tb]
		\includegraphics[width=\linewidth]{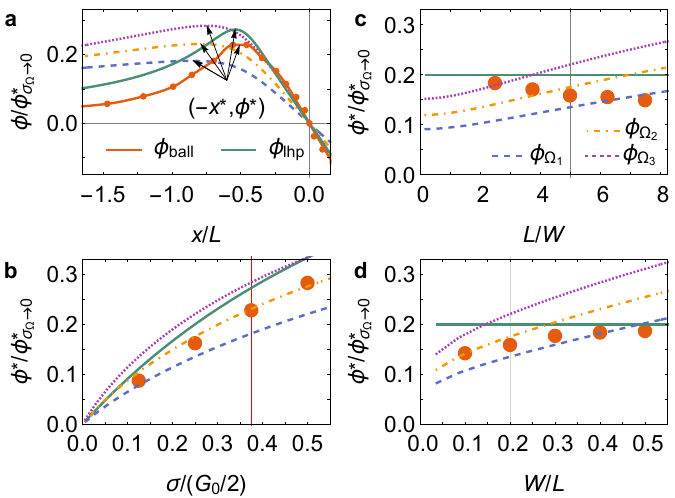}
		\caption{\textbf{Normalized interface potential $\phi_\mathrm{ball}$ of a 1D ballistic edge channel} (solid orange line and orange circles), in comparison with several resistive proxies: quasi-1D Ohmic channels $\Omega_1$ with $\sigma_{\Omega_1}= (G_0/2) L/d = 3874\,\textnormal{\textmu} {S}\square^{-1}$ (dashed blue), $\Omega_2$ with $\sigma_{\Omega_2}= 0.7\,\sigma_{\Omega_1}$ (dash-dotted yellow), $\Omega_3$ with $\sigma_{\Omega_3}= 0.5 \,\sigma_{\Omega_1}$ (dotted purple), and Ohmic lower half-plane with $\sigma_\mathrm{lhp} = G_0/2$ (solid green), \textbf{(a)} as a function of $x$ with $L = 1\,\textnormal{\textmu} \mathrm{m}$, $W = 0.2\,\textnormal{\textmu} \mathrm{m}$, $\sigma = 0.375\,G_0/2= 14.53\,\textnormal{\textmu} {S}\square^{-1}$ in all cases. Note that for simplicity we show only one quadrant, since $\phi(-x)= -\phi(x)$. \textbf{(b)}-\textbf{(d)} The maximum as a function of \textbf{(b)} the sheet conductivity $\sigma$ in the 2D half-plane ($L = 1\,\textnormal{\textmu} \mathrm{m}$, $W= 0.2\,\textnormal{\textmu} \mathrm{m}$), \textbf{(c)} source-to-drain distance $L$ ($\sigma = 0.25 \, G_0/2= 9.69\,\textnormal{\textmu} {S}\square^{-1}$, $W = 0.2\,\textnormal{\textmu} \mathrm{m}$), and \textbf{(d)} distance from the ballistic channel $W$ ($\sigma = 0.25 \, G_0/2$, $L = 1\,\textnormal{\textmu} \mathrm{m}$). The widths of the quasi-1D Ohmic channels are $d=0.01\,\textnormal{\textmu} \mathrm{m}$.
		}
		\label{fig:2}
	\end{figure}

    \textit{Filling condition}.---To retain a perfectly quantized ballistic channel in our setup, all right moving states in the energy interval $[-e\phi^\ast, +e\phi^\ast]$ around the equilibrium chemical potential must be completely occupied and propagate with perfect transmission between the regions of distributed injection and extraction, while obeying the local limits regarding the injection or extraction energies as discussed in the previous section. Implementing these requirements by equating the local current density entering the ballistic channel from the 2D half-plane to the complete filling of its states over all energies that are locally accessible, we obtain the \emph{filling condition}
	\begin{equation} \label{eq:filling-condition}
		\left. \sigma \partial_y\Phi(x,y) \right|_{y=0^{+}} = \frac{G_0}{2} \int_{-\phi(x)}^{\phi(x)}d\phi \, \frac{1}{\Delta x(\phi)}.
	\end{equation}
    The right-hand side is a Riemann-Stieltjes integral from $-\phi(x)$ to the local potential $\phi(x)$ (vertical purple stripes in Figs.~\ref{fig:1}d,f). Note that the equation is nonlocal, as the injection to completely fill the density of states of the ballistic channel at each energy $e \phi$ is distributed over the length $\Delta x (\phi)$ (horizontal red stripe in Fig.~\ref{fig:1}d). The interface potential $\phi(x)$ together with the boundary conditions for the contacts uniquely determine the potential landscape $\Phi(x,y)$ over the complete half-plane via analytical continuation and thereby also fix the injected current density on the left-hand side of Eq.~\eqref{eq:filling-condition}. Examples for the numerically calculated $\Phi(x,y)$ and $\phi(x)$ are shown in Fig.~\ref{fig:1}c and Fig.~\ref{fig:2}a, respectively. The derivation of Eq.~\eqref{eq:filling-condition} and the numerical method for solving it are provided in \ref{sec:derivation} and \ref{sec:Numericalsolution}.
    We stress that the filling condition follows from general considerations (current continuity, energy conservation, symmetry) within semiclassical transport theory applied to the (ideally insulating) half-plane in the presence of residual charge carriers and ballistic transport without scattering in the edge channel~\cite{Leumer2024}, thereby avoiding a self-consistent Schr\"odinger-Poisson treatment of electron density and potential in the mixed 1D/2D system~\cite{Armagnat2019, Flor2022}. We further emphasize that $\phi(x)$ from Eq.~\eqref{eq:filling-condition} is expected to peak in the range $1.5$ to $15$\,mV for typical parameters (\ref{sec:appendixD}), well within the resolution of, e.g., STM-based potentiometry and therefore directly accessible in experiment.

	\textit{Comparison to quasi-1D resistive channels}.---
    To evaluate the distinct transport behavior of a ballistic channel, we compare the interface potential $\phi(x)$ as obtained from the filling condition [Eq.~\eqref{eq:filling-condition}] with the interface potential of a quasi-1D Ohmic channel~\cite{Weber2012}. The latter is modeled as a narrow stripe of width $d \ll W,L$, with $y \in [-d, 0]$ and infinite extension along $x$, having a sheet conductivity $\sigma_\Omega$ (1D conductivity $\sigma_\Omega d$); its potential can be obtained analytically (\ref{sec:appendixC}) and has the same generic shape as that of a ballistic channel (Fig.~\ref{fig:2}a). Natural limiting cases are the perfectly conducting Ohmic channel ($\sigma_\Omega \rightarrow \infty$), with vanishing interface potential ($\phi = 0$) and maximal current, and the fully insulating Ohmic channel ($\sigma_\Omega\rightarrow 0$) carrying zero current. $\phi_{\sigma_{\Omega}\rightarrow 0}$ provides a convenient upper limit, since any finite current injection naturally lowers the interface potential towards zero (the opposite limiting case).
	Hence, we normalize all interface potentials with $\phi_{\sigma_{\Omega}\rightarrow 0}^\ast$, the maximum of $\phi_{\sigma_{\Omega}\rightarrow 0}(x)$.
	
	To explore whether the interface potential of a ballistic channel ($\phi_\mathrm{ball}$) can be emulated by an Ohmic channel, we choose the latter's sheet conductivity $\sigma_{\Omega_1}$ to match the quantized conductance $G_0/2$ over the length $L$, i.e., $\sigma_{\Omega_1} \equiv (G_0/2) L/d$. While the interface potentials $\phi_\mathrm{ball}$ and  $\phi_{\Omega_1}$ (Fig.~\ref{fig:2}a) appear qualitatively similar, there are notable differences: $\phi_{\Omega_1}$ exhibits a slower decay at large $|x| > L$, but a gentler slope towards $x=0$, with a crossover of the interface potentials around $|x| \sim L/2$. These differences arise because the voltage drop of an Ohmic channel scales with the path length, while being path length-independent over a ballistic channel. Adjusting the sheet conductivity of the Ohmic channel, some features of $\phi_\mathrm{ball}$ can be recovered more accurately. However, systematic agreement with $\phi_\mathrm{ball}$ along the whole interface cannot be obtained for \textit{any} quasi-1D Ohmic channel.

\begin{figure}[tb]
		\includegraphics[width=\linewidth]{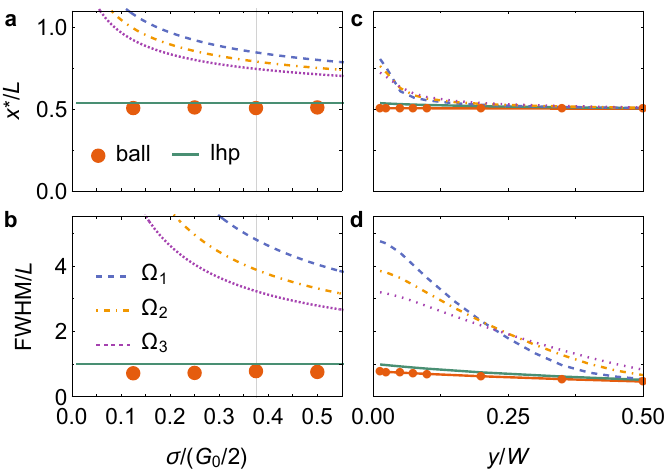}
		\caption{\textbf{(a)},\textbf{(c)} \textbf{Peak position $x^\ast$ of the interface potential and} \textbf{(b)},\textbf{(d)} \textbf{corresponding FWHM}, as a function of \textbf{(a)},\textbf{(b)} the sheet conductivity $\sigma$ in the 2D half-plane and \textbf{(c)},\textbf{(d)} the distance $y$ from the interface ($\sigma = 0.375 \, G_0/2= 14.53\,\textnormal{\textmu} {S}\square^{-1}$), with $L = 1\,\textnormal{\textmu m}$, $W = 0.2\,\textnormal{\textmu m}$ in all cases. For the latter, the potential $\Phi(x,y)$ was considered with $y>0$ fixed, instead of $\phi(x)$.
        Numerical solutions for the ballistic channel are shown by orange circles, analytic solutions of quasi-1D Ohmic channels (see Fig.~\ref{fig:2} for conductivities) $\Omega_1$ (dashed blue), $\Omega_2$ (dash-dotted yellow), and $\Omega_3$ (dotted purple), and a conducting lower half-plane (lhp) with $\sigma_\mathrm{lhp} = G_0/2$ (solid green) are plotted for comparison.
        }
		\label{fig:3}
	\end{figure}

    A better overall agreement with $\phi_\mathrm{ball}$, in particular for $|x|>L$, can be obtained by considering an Ohmic lower half-plane with sheet conductivity $\sigma_\mathrm{lhp} \equiv G_0/2$ in place of the ballistic channel (green profiles in Figs.~\ref{fig:2} and \ref{fig:3}). This boundary value problem can also be solved analytically~\cite{Leis2022b} (\ref{sec:appendixC}). Because of its 2D geometry, a distribution of current paths develops in the lower half-plane that mitigates the linear scaling of the resistance, the general property of all quasi-1D Ohmic channels that is in direct contrast with a ballistic channel. Thus, although the conducting lower half-plane does not represent a 1D channel geometrically and moreover is Ohmic, it turns out to be a reasonable proxy for a 1D ballistic channel in our contact geometry. With respect to experiments, this raises the question how the transport signature of a truly existing lower half-plane (e.g., the substrate terrace next to the edge of a quantum spin Hall insulator) can be distinguished from a ballistic edge channel. Here we note that, to be a good proxy, the lower half-plane has to have a sheet conductivity close to $G_0/2$, while any physical lower half-plane will have $\sigma_{\mathrm{lhp}} \ll G_0/2$, making it irrelevant for the transport problem. If in doubt, $\sigma_{\mathrm{lhp}}$ can be measured independently to confirm that a transport signature similar to the one displayed by the orange symbols in Figs.~\ref{fig:2} and \ref{fig:3} must be due to a ballistic edge channel.

	\textit{Hallmark behavior of a ballistic channel}.---To pinpoint criteria which allow experimental identification of 1D ballistic transport, we explore some characteristic features of the interface potential $\phi$, keeping the total source-to-drain current fixed.
	Figs.~\ref{fig:2}b and \ref{fig:3}a,b show how the maximum $\phi^\ast$ of the interface potential, its position $x^\ast$, and the full width at half maximum (FWHM) of the potential profile (the latter providing a measure of the decay at large distances) depend on the half-plane's sheet conductivity $\sigma$. The same conductivities of Ohmic channels as in Fig.~\ref{fig:2}a are employed. The scaling of $\phi^\ast$ with $\sigma$ is similar for all scenarios (Fig.~\ref{fig:2}b).
    When analogously varying the source-to-drain distance $L$ (Fig.~\ref{fig:2}c) and common distance to the ballistic channel $W$ (Fig.~\ref{fig:2}d), we find a qualitatively different scaling behavior of $\phi^\ast$ with $L$, increasing (decreasing) with $L$ for the Ohmic (ballistic) channel. This reflects the qualitatively different resistive behavior of Ohmic and ballistic channels when changing the path lengths of conduction (see \ref{sec:Numericalsolution} for a more detailed discussion).
    Moreover, all Ohmic channels display a much broader maximum (Fig.~\ref{fig:3}b) that is significantly displaced from the contact positions ($x=\pm L/2$) to larger $x$ (Fig.~\ref{fig:3}a). While the displacement quickly decays when moving away from the interface (Fig.~\ref{fig:3}c), the larger FWHM of Ohmic channels persists to rather large $y$ (Fig.~\ref{fig:3}d), making the narrow FWHM a relevant and experimentally measurable signature that clearly differentiates ballistic from Ohmic channels.
	
	\textit{Conclusion and outlook}.---We extended the standard Landauer approach to a ballistic channel in direct contact with a conducting half-plane to which source and drain contacts are applied. In this geometry, the current enters and leaves the ballistic channel via the half-plane in a spatially distributed manner. The potential at the interface between half-plane and ballistic channel displays distinct features compared to a quasi-1D Ohmic channel, such as a narrower maximum and a steeper decay at large distances, and opposite scaling of the maximum with the distance between the contacts. These features reflect the fundamental ballistic nature of the channel and can be used to identify ballistic channels in various topological materials experimentally, using potentiometry in multi-tip STM, for example. Our results thus widen the scope of multi-probe transport experiments to ascertain the ballistic nature of edge channels.
    
    Our generalization of Landauer's theory to distributed injection can be applied to a wide range of contact geometries in a straightforward manner, by modifying the potential $\Phi(x,y)$ on the left-hand side of Eq.~\eqref{eq:filling-condition} to match the appropriate boundary conditions in the 2D bulk region.
    The right-hand side of Eq.~\eqref{eq:filling-condition} can generally be applied for setups with a symmetric positioning for source and drain, for which $\Delta x (\phi)$ follows naturally from symmetry considerations (\ref{sec:derivation}), while a generalization for asymmetric setups remains for future work.
    Furthermore, the filling condition for distributed injection can be extended to lithographic contacts that cover both 2D bulk and edge~\cite{Wu2018}, by including the current that is directly injected and extracted via the contacts in addition to the distributed one when matching the density of states of the ballistic channel over the appropriate energy range.
    Thus, treating 1D ballistic edge channels in terms of the filling condition paves the way for a wide range of experimentally relevant setups with distributed injection, providing a valuable tool for the realistic treatment of transport in mixed-dimensional (2D/1D or even 3D/1D) systems~\cite{Baringhaus2014, Wu2018, deng2020quantum}.
    
	\textit{Acknowledgments}.---
    The authors would like to thank Arthur Leis for insightful discussions. This work was supported by the German excellence cluster ML4Q (Matter and Light for Quantum Computing) and by the QuantERA grant MAGMA (by the German Research Foundation under grant 491798118). K.M., F.L.\ and F.S.T.\ acknowledge the financial support by the Bavarian Ministry of Economic Affairs, Regional Development and Energy within Bavaria’s High-Tech Agenda Project ``Bausteine f\"ur das Quantencomputing auf Basis topologischer Materialien mit experimentellen und theoretischen Ans\"atzen'' (Grant No.\ 07 02/686 58/1/21 1/22 2/23). K.M.\ acknowledges financial support by the German Federal Ministry of Education and Research (BMBF) via the Quantum Future project ‘MajoranaChips’ (Grant No.\ 13N15264) within the funding program Photonic Research Germany. C.W.\ acknowledges funding through the European Research Council (ERC-StG 757634 “CM3”). F.L.\ acknowledges funding by the Deutsche Forschungsgemeinschaft (DFG, German Research Foundation) within the Priority Programme SPP 2244 (Project No.\ 443416235), as well as the Emmy Noether Programme (Project No.\ 511561801). F.S.T.\ acknowledges funding by the Deutsche Forschungsgemeinschaft (DFG, German Research Foundation) through Coordinated Research Center CRC 1083, project ID 223848855.


    \putbib[bu1.bbl]
    \end{bibunit}
    
	\clearpage
	\widetext
	
	\setcounter{section}{0}
	\setcounter{equation}{0}
	\setcounter{figure}{0}
	\setcounter{table}{0}
	\setcounter{page}{1}
	\makeatletter
	\renewcommand{\thesection}{SUPPLEMENTAL NOTE \arabic{section}}
	\renewcommand{\thesubsection}{\Alph{subsection}}
	\renewcommand{\theequation}{S\arabic{equation}}
	\renewcommand{\thefigure}{S\arabic{figure}}
	\renewcommand{\figurename}{Supplemental Figure}
	\renewcommand{\bibnumfmt}[1]{[S#1]}
	\renewcommand{\citenumfont}[1]{S#1}

    \begin{bibunit}[]

    \begin{center}
		\textbf{Supplemental Material: Distributed Current Injection into a One-Dimensional Ballistic Edge Channel} \linebreak \linebreak
		Kristof Moors,$^{1,2}$ Christian Wagner,$^{3}$ Helmut Soltner,$^{4}$ Felix L\"upke,$^{2,3,5}$ F.~Stefan Tautz,$^{3,2,6}$ Bert Voigtl\"ander$^{3,2,6}$\linebreak\linebreak
		\textit{$^{1}$ Peter Gr\"unberg Institute (PGI-9), Forschungszentrum J\"ulich, 52425 J\"ulich, Germany}\\
		\textit{$^{2}$ Jülich Aachen Research Alliance (JARA), Fundamentals of Future Information Technology, 52425 Jülich, Germany}\\
		\textit{$^{3}$ Peter Gr\"unberg Institute (PGI-3), Forschungszentrum J\"ulich, 52425 J\"ulich, Germany}\\
		\textit{$^{4}$ Institute of Technology and Engineering (ITE), Forschungszentrum J\"ulich, 52425 J\"ulich, Germany}\\
		\textit{$^{5}$ Institute of Physics II, Universit\"at zu K\"oln, Z\"ulpicher Straße 77, 50937 K\"oln, Germany}\\
		\textit{$^{6}$ Experimental Physics IV A, RWTH Aachen University, Otto-Blumenthal-Straße, 52074 Aachen, Germany}\\
		
	\end{center}
    
	\section{Generic shape of the interface potential}
	\label{sec:appendixA}
	
	For the experimental geometry in Fig.~\ref{fig:1}b of the Main Text, which has a mirror plane in the $yz$ plane, symmetry requires that $\phi(x)=-\phi(-x)$ (symmetrically applied bias voltage $\Phi_{S}=-\Phi_{D}$). Furthermore, $\lim_{x\rightarrow\pm\infty}\phi(x)=0$ (potential approaches zero at infinity) and $\lim_{x\rightarrow\pm\infty}\partial_y\Phi(x,y)|_{y=0^+}=0$ (no injection/extraction at infinity). The symmetry $\phi(x)=-\phi(-x)$ holds if and only if the current paths in the 2D half-plane are mirror symmetric at the $yz$ plane, except for their opposite directionality. 
	
	Because there is only a single source and a single drain contact, it is reasonable to assume that $\phi(x)$ reaches a single global maximum $\phi^{\ast}(-x^{\ast})$ and a single global minimum $-\phi^{\ast}(+x^{\ast})$ in the interval $[0,+\infty[$, with $x^\ast >0$. If $\Phi_{S}>\Phi_{D}$, $\phi(x)$ monotonically increases between $x>-\infty$ and $x=-x^{\ast}$, monotonically decreases between $x=-x^{\ast}$ and $x=+x^{\ast}$, and monotonically increases between $x=+x^{\ast}$ and $x<+\infty$. $x^{\ast}$ roughly marks the symmetric positions of the current injecting electrodes. The solid orange and blue lines in Figure~\ref{fig:1}d of the Main Text show the shape of the interface potential with the generic characteristics outlined above.

	\section{Filling condition}
	\label{sec:appendixB}
	
	\subsection{Derivation}
    \label{sec:derivation}
	
	In the following, we determine the current density $dI_\mathrm{ch}(x)/dx$ that must locally be injected into the ballistic channel in order to fill its states over the accessible energy range completely. The resultant expression forms the right-hand side of the \emph{filling condition} in Eq.~\eqref{eq:filling-condition} of the Main Text. To avoid overburdening the explanation, we consider here a positive voltage $V=\Phi_{S}-\Phi_{D}>0$, with the charge current being carried by positive charge carriers moving from source to drain.
	
	According to Landauer, the current through a 1D ballistic conductor with perfect transmission from source to drain is given by $I=(2e^2/h) M (\mu_{S}-\mu_{D})/e$ \cite{Datta1995}, where $\mu_{S}$ and $\mu_{D}$ are the chemical potentials of source and drain, respectively, and $(\mu_{S}-\mu_{D})/e$ is the applied voltage $V$. The factor 2 in the Landauer expression derives from spin, while $M$ is the number of transverse modes in the ballistic conductor. For a single, completely filled ballistic channel (with a quantized conductance $G_0/2=e^2/h$ per spin) the current per unit energy and per spin is given by 
	\begin{equation}
		\label{eq:LandauertransportDOS}
		\frac{dI_\mathrm{ch}(E)}{dE} = \frac{e}{h}.
	\end{equation}
	Here, the term `single channel' refers to the contribution of a single transverse mode of the ballistic conductor. Because all carriers in the ballistic conductor are transported with perfect transmission, $e/h$ plays the role of a `density of (transport) states’ in the ballistic conductor.
	
	In our experimental geometry, carriers are initially injected from the source contacts into the 2D half-plane. Since we assume the source contact and the 2D half-plane to be degenerate, \textit{all} carriers are injected into the 2D half-plane with an energy $E_{S}$ corresponding to the chemical potential $\mu_{S}$ of the source contact, $E_{S}=\mu_{S}$ (see \ref{sec:Analysisfillingcondition}). Carriers then move resistively to the ballistic channel, for example to position $(x<0, y\approx\lambda_\mathrm{mfp})$. Here, $\lambda_\mathrm{mfp}$ is the carrier mean free path in the 2D half-plane. On this path, the carriers lower their energy by $\mu_{S}-e\phi(x)$ [assuming that $\Phi(x,y=\lambda_\mathrm{mfp})\approx \Phi(x,y=0^{+})=\phi(x)$] through Ohmic dissipation. Thus, they arrive at $(x,y\approx\lambda_\mathrm{mfp})$ with an energy $E(x)=\mu(x)=e\phi(x)$. Once carriers enter the thin strip of width $\sim\lambda_\mathrm{mfp}$ next to the ballistic channel, carrier transport may involve additional (i.e., non-Ohmic) dissipative relaxation processes, which are discussed in detail in \ref{sec:Analysisfillingcondition}.
	
    We now consider injection into the ballistic channel. For a given $x<0$, injection is limited to energies $E(x)\leq e\phi(x)\equiv E^\mathrm{in}_\mathrm{up}(x)$ at the ballistic channel. Carriers at the local chemical potential $\mu(x)$, i.e., those arriving at the ballistic channel at $(x, y=0)$ after Ohmic transport on their \textit{entire} path across the 2D half-plane (including the strip), are injected at $E^\mathrm{in}_\mathrm{up}(x)$. Carriers that arrive at $(x,y\approx\lambda_\mathrm{mfp})$ at the local chemical potential and then dissipate additional energy in the strip are injected at $E<E^\mathrm{in}_\mathrm{up}(x)$. Conversely, for a particular positive carrier energy $E$, all positions $x$ with $E \leq e\phi(x)$ form an interval with length $\Delta x(E)$ in which injection can occur. This interval is indicated by the horizontal red stripe in the potential diagram in Fig.~\ref{fig:1}d of the Main Text for a specific example value of $E=e\phi$. At the edges of this interval $\Delta x$, carriers arrive with arrival energy $E(x)=E$ at the ballistic channel $(x,y=0)$, having dissipated the energy $\mu_{S} -e\phi(x)$ during their Ohmic transport through the 2D half-plane (without additional relaxation in the thin strip of width $\lambda_\mathrm{mfp}$), while, at $-x^\ast$, carriers arrive at $(-x^\ast,y\approx \lambda_\mathrm{mfp})$ with energy $\mu(-x^\ast)=e\phi^\ast$ (having dissipated $\mu_{S} -e\phi^\ast$ on their Ohmic path) and then additionally dissipate the energy $e\phi^\ast -E$ in the strip before entering the ballistic channel with energy $E$.  
	
	These considerations allow us to formulate the condition for a complete filling of the ballistic channel: at every energy $E$, the 2D half-plane needs to provide in a spatial interval $\Delta x(E)$ at the interface to the ballistic channel precisely the number of carriers such that a current per unit energy $e/h$ is injected from this spatial interval into the ballistic channel. If this \textit{filling condition} is fulfilled, then the transport density of states of the ballistic channel at this energy is completely filled with charge carriers. A physical argument why the 2D half-plane is always able to provide enough carriers to fill the ballistic channel completely is given in \ref{sec:Analysisfillingcondition}.     
	
\begin{figure}[b]
		\includegraphics[width=0.65\linewidth]{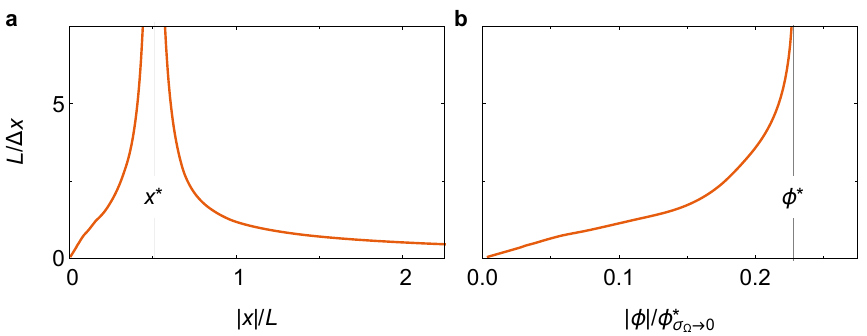}
		\caption{
			The inverse width $1/\Delta x$ \textbf{(a)} as a function of $x$ and \textbf{(b)} as a function of $\phi$ for the interface potential of a ballistic channel that is presented in Fig.~\ref{fig:2} of the Main Text, i.e., for a setup with $L=1\,$\textmu m, $W=0.2\,$\textmu m, and $\sigma=0.375 \, G_0/2$.
		}
		\label{fig:S1}
	\end{figure}

	A few points are worth noting. For the generic shape of the interface potential as discussed in~\ref{sec:appendixA}, we find $\Delta x(E=0) = +\infty$ and $\Delta x (E=e\phi^\ast) = 0$, with $\Delta x(\phi)$ monotonically decreasing between $\phi=0$ and $\phi=\phi^\ast$ (see example in Fig.~\ref{fig:S1}). Furthermore, from the condition $E \leq e\phi(x)$ derived above it would seem that for $E<0$ the spatial injection interval extends from negative infinity to some value $x>0$. However, this is in contrast to Fig.~\ref{fig:1}d of the Main Text, where the interval $\Delta x(E)$ is finite for $E<0$, and in particular ends at a value $x<0$. This additional restriction of the injection interval follows from energy conservation on the extraction side, carrier conservation, and a symmetry argument. The argument is briefly sketched out in the Main Text, but for the sake of completeness it is repeated here. First, we consider energy conservation during extraction of the carriers from the ballistic channel to the 2D half-plane and their subsequent transport to the drain contact. If a carrier is extracted at $x>0$, it must at least have an energy $e\phi(x)$ to enter the 2D half-plane. On its Ohmic path to the drain it will then lower its energy by $e\phi(x)-\mu_{D}$ and enter the drain electrode at its Fermi level $\mu_{D}$. However, it is clear that carriers leaving the ballistic channel with energies $E(x)>e\phi(x)$, can, after loosing $E-e\phi(x)$ through inelastic relaxation processes down to the local chemical potential $\mu(x)=e\phi(x)$, also continue on the Ohmic transport path to the drain, dissipating $e\phi(x)-\mu_{D}$ on their way, and enter the drain at its chemical potential $\mu_{D}$. As on the source side, the inelastic relaxation processes occur in a thin strip of width $\lambda_\mathrm{mfp}$ next to the interface to the ballistic channel. Thus, we conclude that at $x>0$ all carriers with $E\geq e\phi(x)\equiv E^\mathrm{ex}_\mathrm{lo}$ can leave the ballistic channel and reach the drain. 
	
	The principle of energy conservation during the two-stage injection (source $\rightarrow$ 2D half-plane $\rightarrow$ ballistic channel) and extraction (ballistic channel $\rightarrow$ 2D half-plane $\rightarrow$ drain) processes has allowed us to derive the $x$-dependent upper limit for injection and lower limit for extraction. In contrast, the upper limit for extraction and the lower limit for injection follow from the conservation of the number of carriers in the ballistic channel separately at each energy, and symmetry of the current-path distribution in the 2D half-plane. Since in the ballistic channel carriers are transported without loss of energy, the energy distribution of injected carriers must be exactly identical to the energy distribution of extracted carriers. Moreover, the symmetry of the current paths distribution in the half-plane, at the same time a direct consequence and the cause of the antisymmetry of $\phi(x)$, requires that at each $x>0$ precisely the same number of carriers are extracted as injected at the corresponding $-x<0$. Since by construction the energy distribution in the ballistic channel is flat (constant transport density of states $e/h$), this can only be fulfilled for equally wide injection and extraction windows at each pair $-x, x$. 
    Thus, we must conclude
	\begin{equation}
		\begin{split}
			\text{for}\,\, x>0&: \quad E^\mathrm{in}_\mathrm{lo}(-x)=E^\mathrm{ex}_\mathrm{lo}(x)\\
			\text{for}\,\, x<0&: \quad E^\mathrm{ex}_\mathrm{up}(-x)=E^\mathrm{in}_\mathrm{up}(x),
		\end{split}
	\end{equation}
	From this, it follows 
    \begin{equation}
		\begin{split}
			\text{for}\,\, x<0: \quad E^\mathrm{in}_\mathrm{up}(x)&=\phi(x)\\
			&=-\phi(-x)\\
			&=-E^\mathrm{ex}_\mathrm{lo}(-x)\\
			&=-E^\mathrm{in}_\mathrm{lo}(x)\\
			\text{for}\,\, x>0: \quad E^\mathrm{ex}_\mathrm{lo}
			(x)&=\phi(x)\\
			&=-\phi(-x)\\
			&=-E^\mathrm{in}_\mathrm{up}(-x)\\
			&=-E^\mathrm{ex}_\mathrm{up}(x)
		\end{split}
	\end{equation}
	In summary, we arrive at the conclusion that (1) both $E^\mathrm{in}_\mathrm{up}(x)$ and  $E^\mathrm{ex}_\mathrm{lo}(x)$ are located on the $\phi(x)$ curve, (2) both $E^\mathrm{in}_\mathrm{lo}(x)$ and $E^\mathrm{ex}_\mathrm{up}(x)$ are located on the $-\phi(x)$ curve, (3) the injection window at a given $x<0$ equals the extraction window at the corresponding $-x>0$, and (4) at each $x$ the injection and extraction windows are symmetrically arranged around the equilibrium chemical potential in the ballistic channel (which we have defined as the energy zero). These results are displayed in Figs.~\ref{fig:1}d and f of the Main Text.
	
	Now we proceed to derive an explicit mathematical expression for the \emph{filling condition}. By assuming that injection (and thus also extraction) is uniformly distributed over the allowed $\Delta x(E)$, i.e., all allowed positions contribute equally to the total filling of the density of states of the ballistic channel at any given energy, we arrive at the divergence of the channel current per unit energy (see Eq.~\eqref{eq:LandauertransportDOS} and Fig.~\ref{fig:1}d of the Main Text) 
	\begin{equation}
		\frac{\partial^{2}I_\mathrm{ch}(E,x)}{\partial E \partial x}
		= \begin{cases}  \frac{e}{h}\frac{\mathrm{sign}[\phi(x)]}{\Delta x(E)} & |E| \leq e|\phi(x)| \\
			0 & |E| > e|\phi(x)|.
		\end{cases}
		\label{Eq:DOS04}
	\end{equation} 
	This quantity is non-zero if the green square, indicating the point $(E,x)$ at which it is evaluated, is located within the shaded areas in Fig.~\ref{fig:1}d of the Main Text; it is positive in the orange region, and negative in the blue one. The symmetry of our experimental geometry is reflected by
	\begin{equation}
		\begin{split}
			\frac{\partial^{2}I_\mathrm{ch}(E,x)}{\partial E \partial x} &= \frac{\partial^{2}I_\mathrm{ch}(-E,x)}{\partial E \partial x}\\
			\frac{\partial^{2}I_\mathrm{ch}(E,x)}{\partial E \partial x} &= -\frac{\partial^{2}I_\mathrm{ch}(E,-x)}{\partial E \partial x}.
		\end{split}
	\end{equation} 
	The validity of these equations can be easily verified with the help of Fig.~\ref{fig:1}d of the Main Text. The injected current density can now be obtained by integrating over the locally accessible window of possible injection energies, yielding the Riemann-Stieltjes integral
	\begin{equation}
		\begin{split}
			\frac{dI_\mathrm{ch}(x)}{dx}&= \int\limits_{-\infty}^{+\infty} \frac{\partial^{2}I_\mathrm{ch}(E,x)}{\partial E\partial x} \, dE
			= \frac{e}{h}\int_{-e |\phi(x)|}^{+e |\phi(x)|}  
			\frac{\mathrm{sign}[\phi(x)]}{\Delta x(E)} \, dE\\
			&= G_0
			\int_{0}^{\phi(x)}
			\frac{1}{\Delta x(\phi)}  \,d\phi =\frac{G_0}{2}
			\int_{-\phi(x)}^{\phi(x)}
			\frac{1}{\Delta x(\phi)}  \,d\phi.
			\label{Eq:integral1}
		\end{split}   
	\end{equation}
	The last equation follows from symmetry properties (Fig.~\ref{fig:1}d of the Main Text). The integration range of the final expression corresponds to the vertical purple stripe in Fig.~\ref{fig:1}d of the Main Text, plotted for an arbitrary example position $x$. 
	
	Because of current conservation, the \emph{divergence} $dI_\mathrm{ch}/dx$ of the channel current must equal the \textit{density} $dI_\mathrm{inj}/dx$ of the injection current, the direction of $I_\mathrm{inj}$ being defined such that injection of carriers into the channel yields $I_\mathrm{inj}>0$. The  current density in the 2D half-plane  is given by $\mathbf{j} = \sigma \mathbf{E} = - \sigma \nabla \Phi(x,y)$, from which the injection current density follows as 
	\begin{equation} \label{Eq:didx}
		\frac{dI_\mathrm{inj}(x)}{dx}=-j_y= \left. \sigma \partial_y\Phi(x,y) \right|_{y=0^{+}}.
	\end{equation}
	The minus sign in the first equation is a consequence of our definition of the injection current, which in our experimental geometry is directed in the negative $y$-direction, to be positive, while according the general equation for $\mathbf{j}$ a current density $j_y$ of positive carriers in this direction is negative.
	Equating Eq.~\eqref{Eq:integral1} to  Eq.~\eqref{Eq:didx}, we obtain the \emph{filling condition} that is presented in Eq.~\eqref{eq:filling-condition} of the Main Text, repeated here for convenience: 
	\begin{equation*} 
		\left. \sigma \partial_y\Phi(x,y) \right|_{y=0^{+}} = \frac{G_0}{2} \int_{-\phi(x)}^{\phi(x)}\ \frac{1}{\Delta x(\phi)} \,d\phi.
	\end{equation*}
	
	For solving the filling condition, we apply the superposition principle and split the solution into two parts,
	\begin{equation} \label{Eq:sum}
		\Phi(x,y) = \Phi_{{SD}}(x,y) + \Phi_{\text{ball}}(x,y),
	\end{equation}
	where $\Phi_{SD}(x,y)$ fulfills the Poisson equation $\Delta \Phi_{SD}=-\nabla \cdot \mathbf{j}/\sigma$ with the boundary condition $\Phi_{SD}(x,y=0^+)=0$, and $\Phi_\mathrm{ball}(x,y)$ fulfills the Poisson equation $\Delta \Phi_\mathrm{ball}=0$ with boundary condition $\Phi_\mathrm{ball}(x,y=0^+)\equiv\phi(x)$. 
	$\mathbf{j}(x,y)$ is the current density in the 2D half-plane, and its divergence is given by injection at the source and extraction at the drain contacts, $\nabla \cdot \mathbf{j} =I_{SD} [\delta(\mathbf{r}-\mathbf{r}_{S})-\delta(\mathbf{r}-\mathbf{r}_{D})]$. Summing these two equations, we obtain the Poisson equation for $\Phi(x,y)$ with the correct boundary conditions for our experimental geometry.
	
	The solution of the Poisson equation for $ \Phi_{SD}$ yields the analytical expression~\cite{Leis2022b} (see also Eqs.~\eqref{eq:Phi_I_final_sigma2_infinite} and \eqref{eq:Phi_I_final_d=0_sigma3_infinite} in \ref{sec:appendixC})
	\begin{equation}
		\label{eq:Phi_SD}
		\Phi_{SD}(x,y) = \frac{I_{SD}}{4 \pi \sigma} \left[ \ln \left( \frac{(x-L/2)^2 + (y-W)^2}{(x+L/2)^2 + (y-W)^2} \right) - \ln \left( \frac{(x-L/2)^2 + (y+W)^2}{(x+L/2)^2 + (y+W)^2} \right) \right].
	\end{equation}
	In this equation, the current $I_{SD}$ corresponds to the current that is injected via the source contact,
	\begin{equation}
		\label{eq:I_SD}
		I_{SD} = \int\limits_A \mathbf{\nabla}\cdot \mathbf{j}\,dA=\oint\limits_P \mathbf{j}\cdot \mathbf{n}\,dP=-\sigma\oint\limits_P \mathbf{\nabla} \Phi_{SD}(x,y) \cdot \mathbf{n}\, dP
	\end{equation}
	for an area $A$ covering the source contact and a closed path $P$ that encloses it. Evidently, Eqs.~\eqref{eq:Phi_SD} and \eqref{eq:I_SD} are independent of the solution $\Phi_\mathrm{ball}(x,y)$.
	
	In contrast to $\Phi_{SD}(x,y)$, for the potential $\Phi_\mathrm{ball}(x,y)$ there is no analytical expression. Rather, we have to find a solution numerically (see \ref{sec:Numericalsolution} below for details), such that the total potential $\Phi(x,y)$ satisfies the filling condition (Eq.~\eqref{eq:filling-condition} of the Main Text). The numerical solution can be restricted to finding $\phi(x)\equiv \Phi(x,y=0^+) = \Phi_\mathrm{ball}(x, y=0^+)$ (since $\Phi_{SD}(x,y=0^+)=0$ by construction), because $\Phi_\mathrm{ball}(x,y)$ can be related to $\phi(x)$ through harmonic extension of $\phi(x)$ in the half-plane ($y>0$)~\cite{Prosperetti2011}, yielding
	\begin{equation}
		\Phi_\mathrm{ball}(x,y) = \frac{1}{\pi} \int\limits_{-\infty}^{+\infty}  \frac{y}{(x'-x)^2 + y^2} \phi(x')\, dx'.
	\end{equation}
	With this, we can express the left-hand side of Eq.~\eqref{eq:filling-condition} in the Main Text purely in terms of the unknown function $\phi(x)$, as is already the case for the right-hand side (note that the expression for $\partial_y \Phi_{SD}(x,y)|_{y=0^+}$ is derived in \ref{sec:appendixC}, see Eq.~\eqref{eq:injection_current_pc}):
	\begin{align}
		\left. \partial_y \Phi(x,y)\right|_{y=0+} &= \left. \partial_y \Phi_{SD}(x,y)\right|_{y=0^+} + \left. \partial_y \Phi_\mathrm{ball}(x,y)\right|_{y=0^+}, \\
		\left. \partial_y \Phi_{SD}(x,y)\right|_{y=0^+} &= -\frac{I_{SD}}{4 \pi \sigma} \frac{8 L W x}{[(x-L/2)^2 + W^2][(x+L/2)^2 + W^2]}, \\ \label{eq:Hilbert}
		\left. \partial_y \Phi_\mathrm{ball}(x,y)\right|_{y=0^+} &= -\mathcal{H}[d \phi(x)/dx](x).
	\end{align}
	Here, $\mathcal{H}$ denotes the Hilbert transform, which is defined via the principal-value integral
	\begin{equation}
		\mathcal{H}[f(x)](x) \equiv \frac{1}{\pi} \mathrm{p.v.} \int\limits_{-\infty}^{+\infty} \frac{1}{x - x'} f(x')\, d x' .
	\end{equation}
	Eq.~\eqref{eq:Hilbert} follows from 
	\begin{equation}
		\begin{split}
			\left. \frac{\partial \Phi_\mathrm{ball}(x,y)}{\partial y} \right|_{y=0^+} &= \lim_{y \rightarrow 0^+} \lim_{h \rightarrow 0} \frac{1}{\pi} \int\limits_{-\infty}^{+\infty} \frac{1}{h} \left( \frac{y+h}{(x'-x)^2 + (y+h)^2} - \frac{y}{(x'-x)^2 + y^2} \right) \phi(x')\, dx' \\
			&= - \lim_{y \rightarrow 0^+} \lim_{h \rightarrow 0} \frac{1}{\pi} \int\limits_{-\infty}^{+\infty} \frac{1}{h} \left[ \arctan\left( \frac{x' - x}{y + h} \right) - \arctan\left( \frac{x' - x}{y} \right) \right] \frac{d}{d x'} \phi(x') \, dx' \\
			&= \lim_{y \rightarrow 0^+} \frac{1}{\pi} \int\limits_{-\infty}^{+\infty} \frac{x' - x}{(x' - x)^2 + y^2} \frac{d}{d x'} \phi(x') \, dx' = - \mathcal{H}\left[ \frac{d}{d x} \phi(x) \right](x).
		\end{split}
	\end{equation}
	Also note that the Hilbert transform commutes with the derivative: $\mathcal{H}[d f(x)/d x](x) = d \mathcal{H}[f(x)](x)/d x$.
	
	Thus, the filling condition can finally be written as
	\begin{equation} \label{eq:filling-condition-alt}
		- \frac{I_{SD}}{4 \pi} \frac{8 L W x}{[(x-L/2)^2 + W^2][(x+L/2)^2 + W^2]} - \sigma \mathcal{H}[d\phi(x)/dx](x) = G_0 \int_0^{\phi(x)}\frac{1}{\Delta x(\phi)}\, d \phi .
	\end{equation}
	In the following subsection, we discuss the numerical procedure for retrieving $\phi(x)$ from Eq.~\eqref{eq:filling-condition-alt}.
	
	\subsection{Numerical solution}
	\label{sec:Numericalsolution}
\begin{figure}
		\includegraphics[width=0.8\linewidth]{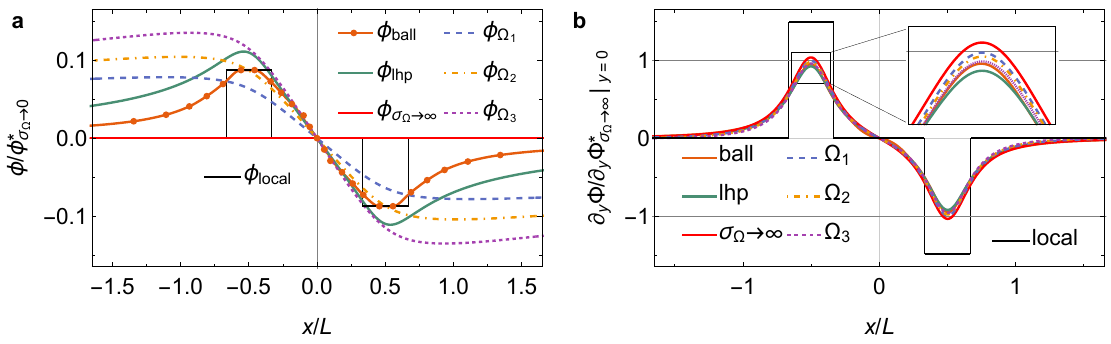}
		\caption{
			(a) The interface potential profile for distributed injection as in Fig.~\ref{fig:2} in the Main Text ($L=1\,$\textmu m, $W=0.2\,$\textmu m), with the sheet conductivity of the half-plane of injection reduced to $\sigma = 0.125 \, G_0/2$ and discrete parametrization indicated.
			(b) The injected current density profiles for the interface potential profiles depicted in (a), normalized by the current density at $x^\ast_{\sigma_\Omega \rightarrow \infty}$ of a perfectly conducting Ohmic channel (see ~\ref{subsubsec:pc}).
			We also show here the profiles for a perfectly conducting proxy ($\phi_{\sigma_\Omega\rightarrow\infty}$) and a standard Landauer treatment of a ballistic channel with local injection ($\phi_\mathrm{local}$), considering source and drain electrodes of width $L/3$ at constant potential $\phi^\ast$ and injection/extraction uniformly distributed over their widths (see text for details).
		}
		\label{fig:S2}
	\end{figure}

	We represent the generic antisymmetric shape of the interface potential by a parameter $\phi^\ast$ (maximum of $\phi$), a sequence of monotonically increasing potential values $\phi_i$ with $0 < \phi_i < \phi^\ast$ and a corresponding monotonically decreasing sequence of position parameters $x_i$ ($x_{i+1} < x_i < 0$) such that $\phi(x_i) = \phi_i$, and another sequence of monotonically increasing potential values $\tilde{\phi}_j$ with $0 < \tilde{\phi}_j \leq \phi^\ast$ and a corresponding monotonically increasing sequence of position parameters $\tilde{x}_j$ ($\tilde{x}_{j-1} < \tilde{x}_j \leq \min_i\{x_i\}$) such that $\phi(\tilde{x}_j) = \tilde{\phi}_j$. Based on this discrete parametrization  $\{ \phi^\ast, \{x_i\}, \{\tilde{x}_j\} \}$ and $\phi(x)$ being an odd function, we can retrieve a continuous interface potential $\phi(x)$ via interpolation.
	For the interpolation between the discrete points, we employ an interpolation scheme based on piecewise cubic Hermite interpolating polynomials (PCHIP), which retains the monotonicity of the profile. At small $x$, we connect the solution to $\phi(x=0)=0$ while, for the decay profile at large $|x|$, we fit a power law to match the power-law decay of the interpolated profile near the most distant position of the sequence, $\min_j\{\tilde{x}_j\}$. In Fig.~\ref{fig:S2}a (orange curve), an example continuous interface potential profile is shown that is obtained from such a discretized parametrization, with a sequence of six points between $x=-x^\ast$ and $x=0$ with $x^\ast = - (\min_i\{x_i\} + \max_j\{\tilde{x}_j\})/2$, and another sequence of six points between $x = -\infty$ and $x = -x^\ast$ (filled orange circles).
	To have well-behaved monotonicity near the maximum, we allow for a flat maximum with a small width (i.e., $|\min_i\{x_i\} - \max_j\{\tilde{x}_j\}| \ll L$) with our parametrization. We find that such a parametrization is sufficient for getting interface potential profiles $\phi(x)$ that satisfy the filling condition reasonably well over a wide range of setup parameters ($L, W, \sigma$).
	
	In detail, to obtain a $\phi(x)$ which fulfills the filling condition, we minimize the difference
	\begin{equation}
		\begin{split}
			\delta I &\equiv \int\limits_{-\infty}^{0} d x \, \left| \frac{d I_\mathrm{inj}(x)}{d x} - \frac{d I_\mathrm{ch}(x)}{d x} \right| \\
			&= \int\limits_{-\infty}^{0} d x \left| \frac{I_{SD}}{4 \pi} \frac{8 L W x}{[(x-L/2)^2 + W^2][(x+L/2)^2 + W^2]} + \sigma \mathcal{H}[d \phi(x)/dx](x) + G_0 \int_0^{\phi(x)} \frac{1}{\Delta x(\phi)} \,d\phi \right| .
		\end{split}
	\end{equation}
	between the left-hand side and the right-hand side of the filling condition Eq.~\eqref{eq:filling-condition-alt}. For the numerical evaluation of the second term on the left-hand side of Eq.~\eqref{eq:filling-condition-alt} (i.e., the contribution of $\Phi_\mathrm{ball}$ to the injected current density), the discretely parametrized $\phi(x)$ (with parameter set $\{ \phi^\ast, \{x_i\}, \{\tilde{x}_j\} \}$) is interpolated to a dense grid between positions $x \ll -L$ and $x=0$; the Hilbert transform is then implemented with discrete (inverse) Fourier transforms. The width function $\Delta x(\phi)$ is also constructed from the  $\phi(x)$ on this dense grid, such that the right-hand side of Eq.~\eqref{eq:filling-condition-alt} can be evaluated numerically. Finally, the integral $\delta I$ is effectively calculated on a dense discrete grid of $x$ values, first for an initial guess parameter set $\{ \phi^\ast, \{x_i\}, \{\tilde{x}_j\} \}$, and then for parameters that are iteratively updated through the Nelder-Mead method to minimize $\delta I$. Through this minimization procedure, we find a solution for the interface potential $\phi(x)$ with which the local injection from the conducting half-plane and the expected local injection for a completely filled ballistic channel are in good agreement. The full numerical procedure is implemented in Python and is openly available~\cite{data}.
	
	After analytic continuation of the interface potential $\phi(x)$ to  $\Phi_\mathrm{ball}(x,y)$, the complete potential $\Phi(x,y)$ can obtained from Eq.~\eqref{Eq:sum}, and thus the injected current density $dI_\mathrm{inj}/dx$ can be calculated using Eq.~\eqref{Eq:didx}. An exemplary result is presented in Fig.~\ref{fig:S2}b together with the injected current densities for the quasi-1D Ohmic channels $\Omega_1$ to $\Omega_3$ and the conductive lower half-plane lhp (all of them obtained from analytical expressions, see~\ref{sec:appendixC}). Note that the displayed interface potential profiles and their characteristics (e.g., $\phi^\ast$, $x^\ast$, FWHM) differ much more among each other than the current density profiles do, which are all rather close to the current density profile of a perfectly conducting edge, $\partial_y \Phi_{SD}(x,y)|_{y=0^+} = \partial_y \Phi_{\sigma_\Omega \rightarrow \infty}|_{y=0^+}$, which can be expected when the upper half-plane is poorly conducting compared to the ballistic channel or Ohmic proxy.
	We can also compare the injection current density profile for distributed injection to the one obtained in a standard Landauer treatment for electrodes that directly inject and extract the channel current locally over channel sections with a finite length $L_\mathrm{el}$. Assuming the same bias window $[-\phi^\ast, +\phi^\ast]$ as for distributed injection and a uniform spatial distribution of the injected (and extracted) current over the length $L_\mathrm{el}$, the injected current density is equal to $d I_\mathrm{inj} / d x = G_0 \phi^\ast / L_\mathrm{el}$. This profile is also shown in Fig.~\ref{fig:S2}b for electrodes that are centered around $\pm x^\ast$ and have a width $L_\mathrm{el} = L/3$.
	
\begin{figure}
		\includegraphics[width=0.85\linewidth]{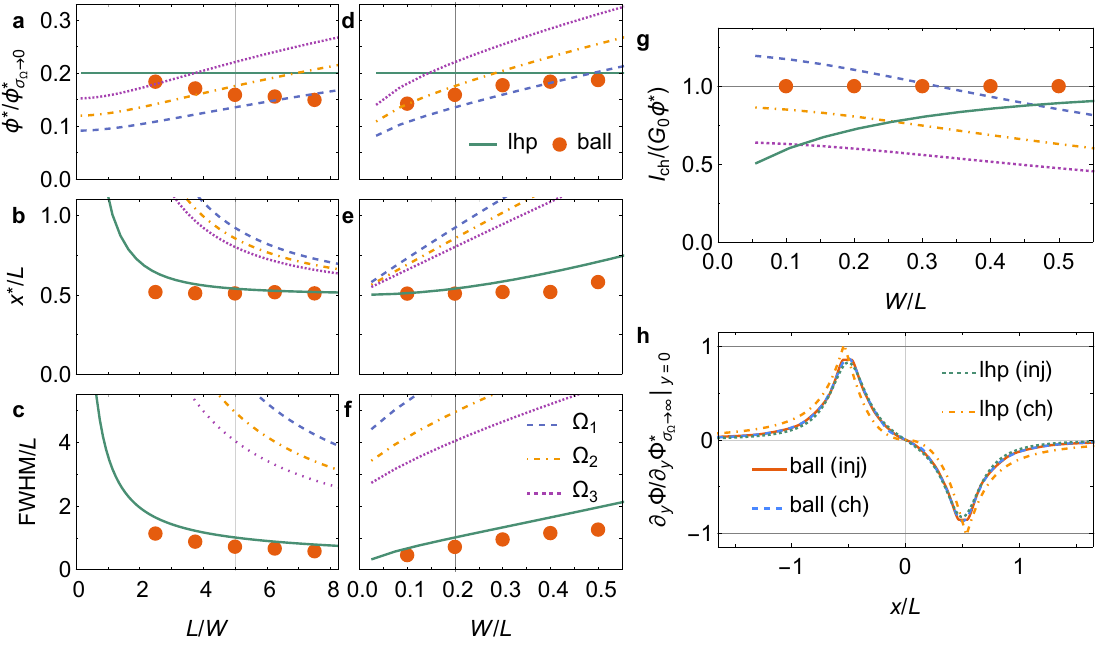}
		\caption{
			(a),(d) Maximum of the interface potential $\phi^\ast$ normalized to $\phi^\ast_{\sigma_{\Omega}\rightarrow 0}$, (b),(e) the corresponding position $x^\ast$, and (c),(f) the full width at half maximum (FWHM) (a)-(c) as a function of the distance between the contacts $L$, considering a common distance $W = 0.2\,$\textmu m of the contacts to the ballistic channel, and (d)-(f) as a function of $W$, considering $L = 1\,$\textmu m.
            (g) The total channel current as a function of $W$, considering $L = 1\,$\textmu m.
            In panels (a) to (g), the numerical solutions for a ballistic channel with the filling condition, Eq.~\eqref{eq:filling-condition} of the Main Text, are shown (orange filled circles labelled ball), together with analytic solutions for different quasi-1D Ohmic channels ($\Omega_\text{1-3}$) and a conducting lower half-plane as proxy (lhp) for comparison, with line styles and conductivities $\sigma_{\Omega_1}$, $\sigma_{\Omega_2}$, $\sigma_{\Omega_3}$, and $\sigma_{\mathrm{lhp}}$ as in Figs.~\ref{fig:2} and \ref{fig:3} of the Main Text.
			(h) The current density (inj) according to the injected current density from the upper half-plane (see Eq.~\eqref{Eq:didx}), and the current density (ch) according to the complete-filling requirement (see Eq.~\eqref{Eq:integral1}), for the ballistic channel (ball) and for a conducting lower half-plane as proxy (lhp), considering $W = 0.2\,$\textmu m and $L = 1\,$\textmu m.
            A sheet conductivity $\sigma = 0.25 \, G_0/2$ for the 2D upper half-plane (from which injection occcurs) is considered in all panels, and the common solution between (a)-(c) and (d)-(f) is indicated by thin gray vertical lines.
            }
		\label{fig:S3}
	\end{figure}

    For the ballistic channel, a numerical evaluation of the  maximum $\phi^\ast/\phi_{\sigma_{\Omega}\rightarrow 0}^\ast$ of the normalized interface potential, its position $x^\ast$, and the full width at half maximum (FWHM) of the potential profile (the latter providing a measure of the decay at large distances)  is presented in Figs.~\ref{fig:S3}a-f as a function of distances $L$ and $W$ and compared to the corresponding solutions for the quasi-1D Ohmic channels $\Omega_1$ to $\Omega_3$ and the conductive lower half-plane lhp. 
    Compared to the ballistic channel, the Ohmic channels display broader maxima, shifted to larger distances (Figs.~\ref{fig:S3}b,c,e,f). As discussed in the Main Text, these differences arise because the voltage drop of an Ohmic channel scales with the path length, while being path length-independent over a ballistic channel.  
    Overall, the trends for both the ballistic case and the Ohmic cases when \emph{increasing} the distance $L$ (Fig.~\ref{fig:S3}a-c) are qualitatively similar to the trends when \emph{decreasing} the distance $W$ (Fig.~\ref{fig:S3}d-f), with the respective other parameters kept fixed. There is, however, one exception: the normalized maximum interface potential $\phi^\ast/\phi^\ast_{\sigma_\Omega\rightarrow 0}$ for the quasi-1D Ohmic channels \emph{increases} as a function of both $W$ and $L$ (Fig.~\ref{fig:S3}a, d), while for the ballistic channel there is an \emph{increase} as a function of $W$ (Fig.~\ref{fig:S3}d) and a \emph{decrease} as a function of $L$ (Fig.~\ref{fig:S3}a). Thus, the scaling trend of $\phi^\ast/\phi^\ast_{\sigma_\Omega\rightarrow 0}$ for the ballistic and quasi-1D Ohmic channels is opposite when changing $L$, but the same when changing $W$. Notably, this is because $L$, unlike $W$, has a direct impact on the path lengths of conduction along the channel and changes the resistance of the (ballistic or Ohmic) edge channel relative to the Ohmic upper half-plane in opposite directions: with $L$, this relative resistance rises for the Ohmic channels, but decreases for the ballistic one. Hence, changing the distance between electrodes is the best approach for discriminating between a ballistic and a quasi-1D Ohmic edge channel through experimental potentiometry in the experimental geometry considered here. Further note that the conventional scenario of local injection is only retrieved in the limit $L \gg W$, whence $x^\ast \rightarrow L/2$ and $\mathrm{FWHM}/L \rightarrow 0$.

	In Fig.~\ref{fig:S3}g, the total channel current that is obtained by integrating the injected current density over the distributed source region $x\in ]-\infty,0[$ is displayed. For the solution of the ballistic channel with complete filling over the bias window $[-\phi^\ast, \phi^\ast]$, the conductance $I_\mathrm{ch}/(2\phi^\ast)$ is systematically quantized to $G_0/2$. This is in general not the case for any Ohmic medium, except when the quasi-1D Ohmic channel conductivity and the geometry parameters $W$ and $L$ are carefully chosen (see $\phi_{\Omega_1}$ with $W \approx 0.3 \, L$), or in the limit $W \gg L$ for the half-plane proxy lhp. Furthermore, the local balance between the injected current density via the Ohmic upper half-plane (Eq.~\eqref{Eq:didx}) and the current density that is required to fill a \textit{ballistic} channel in the bias window (Eq.~\eqref{Eq:integral1}) is only achieved for the interface potential of a ballistic channel that satisfies the filling condition and not by any Ohmic medium as replacement. This is demonstrated in Fig.~\ref{fig:S3}h, where both current densities are evaluated and presented for a numerical solution of the \emph{filling condition}, and for an Ohmic lower half-plane (lhp) proxy.
	
    \subsection{Analysis of the contact resistance, heat dissipation and filling condition for the case of distributed injection}
	\label{sec:Analysisfillingcondition}
	
	Before turning to a deeper analysis of distributed injection in our experimental geometry (Fig.~\ref{fig:1}b of the Main Text), we recap the situation for the standard Landauer problem (Fig.~\ref{fig:1}a of the Main Text). In the Landauer approach, the properties of a 1D ballistic channel follow from a confinement argument. At either contact (source and drain), the $y$ dimension is reduced to a very narrow strip, while the $x$ dimension is left unchanged (see Fig.~\ref{fig:1}a of the Main Text). Consequently, the effectively infinite set of $k_y$ quantum numbers is reduced to a small, discrete number $M$ of  transverse modes in the ballistic channel. In contrast, the number of allowed $k_x$ in the ballistic channel remains large (of similar order as in the contact electrode, if the ballistic channel has macroscopic length). Thus, the current in the contacts can be (and is) carried by infinitely many modes (all having distinct $k_y$ for a given $k_x$), while in the ballistic channel it must be carried by only a few modes---the $M$ quantized sub-bands (transverse modes) that follow from the confinement in $y$ direction. Because there is no momentum conservation in $k_y$ at the contact point between the electrode and the ballistic channel, carriers with any $k_y$ in the electrode can elastically scatter into the few transverse modes in the ballistic channel. Similarly, the lack of momentum conservation in $k_x$ means that carriers that pass elastically from the contact electrode to the ballistic channel can always change their $k_x$ to match the latter's dispersion relation $E(k_x)$. Thus, any right-moving state ($k_x>0$) in the source electrode can transition to the ballistic channel, and conversely, every right-moving state in the ballistic channel can be occupied from the source electrode. Yet, because the number of available modes in the ballistic channel is much smaller than in the source electrode, a bottleneck for carrier transport emerges at the source contact. The well-known consequences are:
	\begin{itemize}
		\item  Since all $M$ modes in the ballistic channel can and will be filled from the source contact, the conductance of the ballistic channel is proportional to $M$, i.e.,~$G=2e^2M/h$.
		\item Because the number of available states in the drain contact is so much larger than in the ballistic channel, it is reasonable to assume that the exit
        from the ballistic channel to the drain contact is reflection-less.
		\item  In the metallic electrodes, a tiny imbalance between quasi-Fermi levels (or chemical potentials) for right- and left-moving carriers (and a corresponding slight asymmetry in the occupation of $k_x>0$ and $k_x<0$ states) is sufficient to support any realistic current, because of the large number of transverse modes (and thus large density of states). In contrast, the difference between quasi-Fermi levels needs to be much larger in the ballistic channel to support the same current.  
		\item Because of their overwhelmingly large number of transverse modes, the contact electrodes pin the chemical potentials in the ballistic channels: right-moving states ($k_x>0$) in the ballistic channel have the same chemical potential as the source electrode, while left-moving states ($k_x<0$) have the same chemical potential as the drain electrode. Indeed, because the exit into the drain contact is reflection-less, causality implies that the source electrode can have no influence on the left-moving states (and in particular their chemical potential), and similarly the drain electrode can have no influence on right-moving states. 
		\item From the difference in the quasi-Fermi levels it follows that the average chemical potential in the ballistic channel is $\mu=(\mu_{S} + \mu_{D})/2$. This implies voltage drops of $V= (\mu_{S} - \mu_{D})/2e$ between source electrode and ballistic channel, as well as the channel and the drain electrode. 
		\item Because the voltage drops at the contacts, the resistance of the ballistic channel, too, has to be located there: At source and drain, the contact resistance is $G_\mathrm{c}^{-1}=h/(4e^2M)$. 
		\item Heat is dissipated where the average chemical potential drops---at the contacts. Fig.~\ref{fig:1}e of the Main Text helps to understand the mechanism of this energy dissipation: Carriers that are injected from the source electrode at $\mu_{S}$ travel elastically to the drain, where they end up in an excited state far above the Fermi level $\mu_{D}$. Inelastic scattering processes in the drain electrode relax the energy of these carriers to the local Fermi level $\mu_{D}$. In contrast, a carrier that is injected at the source from the lower end of the injection window (i.e., at $\mu_{D}$) will enter the drain electrode at its Fermi level. In this case, no energy relaxation in the drain is possible, but it takes place in the source electrode, by filling the carrier hole left behind by the injected carrier with a carrier from $\mu_{S}$ (this process can be understood as the scattering of the remaining hole). In this way, precisely half of the energy is dissipated at each of the contacts---as it must, since the contact resistance at both contacts are the same. In the inelastic collisions of the holes in the source and the carriers in the drain, also their respective momentum distributions are changed from a directed distribution to an essentially isotropic one. This is consistent with the situation in the electrodes, in which the difference between the quasi-Fermi levels for left movers and right movers is tiny, and the current is carried by uncompensated electrons in a small crescent at the Fermi level, created by displacing the Fermi sphere by the drift vector $\mathbf{k}_{D}$ in the direction of the current \cite{Datta1995}.    		
	\end{itemize}
 
    We now analyze the case of distributed injection. In the experimental geometry of Fig.~\ref{fig:1}b of the Main Text, there are two types of interfaces, one between the contact electrodes and the 2D half-plane, and one between the half-plane and the ballistic channel. If the electrodes are not too sharp, we can expect a large number of modes in both the metal contacts (loose confinement in $x,\,y$) and the 2D half-plane (strict confinement in $z$), all of which can scatter into each other, since there is no strict momentum conservation (only loose $k_x,\,k_y$ conservation and no $k_z$ conservation on crossing the interface). Thus, we do not expect a significant contact resistance at this interface. Because the electrodes and the 2D half-plane are Ohmic, the quasi-Fermi levels for `forward' movers (i.e., those carriers with $\mathbf{k}$ that corresponds to movement from source to drain) and `backward' movers ($\mathbf{k}$ corresponding to movement from drain to source) are nearly identical, and the current in both the electrodes and the 2D half-plane is carried by uncompensated electrons in a small crescent in forward direction at the Fermi level \cite{Datta1995}. Even if this difference changes slightly between the electrodes and the 2D half-plane (due to a small residual contact resistance), it will always be negligible compared to the difference between the quasi-Fermi levels for right (or `forward') movers and left (or `backward') movers in the ballistic channel. For this reason, we have drawn the carrier traces in Fig.~\ref{fig:1}f of the Main Text as single lines (green, orange, blue), indicating schematically the nearly identical quasi-Fermi levels which follow the electrostatic potential $\Phi(x,y)$ in the 2D half-plane. However, as also illustrated in Fig.~\ref{fig:1}f of the Main Text, an exception from the standard Ohmic transport in most of the 2D half-plane occurs in the vicinity of the ballistic channel. The discussion of the interface between the 2D half-plane and the ballistic channel in the following paragraphs reveals the reason for this. 

    A requirement for the appearance of the quantized conductance in the Landauer picture of local injection is that the number of transverse modes in the contacts has to be much larger than the number of modes in the ballistic channel (ideally a single mode) \cite{Datta1995}. This requirement is also fulfilled in the case of distributed injection at the interface between the 2D half-plane and the 1D ballistic channel (Fig.~\ref{fig:1}b of the Main Text), because the former supports a much larger number of modes than the latter. In fact, the only difference to the standard Landauer picture of local injection (from two 2D contacts, see Fig.~\ref{fig:1}a of the Main Text) is the relative direction of the confinement in the 1D channel: for standard local injection, confinement and direction of transport \textit{in the contact} are perpendicular to each other, while for distributed injection the confinement occurs parallel to at least a component of the transport direction in the contact, i.e., the 2D half-plane. Because this purely geometrical difference has no influence on mode numbers, it is reasonable to assume that a contact resistance exists at the one-dimensional interface to the ballistic channel in Fig.~\ref{fig:1}b of the Main Text, due to a bottleneck in mode numbers (in full analogy to the situation in Fig.~\ref{fig:1}a of the Main Text), and thus all states of the ballistic channel are filled on the source side and the corresponding carriers exit in a reflection-less manner on the drain side, as in the standard Landauer picture. We note, however, that the special geometry of distributed injection (Fig.~\ref{fig:1}b of the Main Text) requires specific scattering when carriers enter or leave the ballistic channel. This scattering occurs in conjunction with the energy relaxation close to the interface, to which we now turn.   

    As we have argued in the Main text (and in a more detailed manner in \ref{sec:derivation}), the ballistic channel draws carriers in a finite energy range of width $2e|\phi(x)|$ from the Fermi sea of the 2D half-plane, conserving their energy. As a consequence, at position $-x'<0$ holes between the local Fermi-level $\mu(-x')$ and $\mu(-x')-2e|\phi(-x')|$ are generated in the 2D half-plane. These holes and the energy range in which they are created are shown in Fig.~\ref{fig:1}f of the Main Text at position $-x'$ and $-x^\ast$. In the ideal case of a translationally invariant interface, momentum conservation for $k_x$ holds, and therefore the holes in the Fermi sea appear at the same finite values of $k_x$ that the corresponding carriers have in the ballistic channel; even in this ideal case, the $k_y$ value of the hole (since $k_y$ is not conserved) provides the necessary freedom to adjust the carrier's $k_x$ for a given energy $E$ to the $E(k_x)$ dispersion in the ballistic channel. Any deviations from an ideal interface will provide additional scattering that safeguards that carriers with the right energy and momenta to fill the ballistic channel are present.
    For reasons of charge conservation, it is clear that the holes must ultimately be filled from the reservoir of uncompensated charge carriers in the crescent around the quasi-Fermi level at position $(x=-x', y\approx 0)$ (since these carry the current), as indicated by orange shading at $-x^\ast$ and $-x'$ in Fig.~\ref{fig:1}f of the Main Text. Through inelastic processes in the Ohmic medium, they cascade down in energy; during this cascade, the carriers also scatter from the momenta with which they arrive at the interface to the momenta of the holes left behind by the carrier injection into the ballistic channel. Since the medium is Ohmic, it can support this scattering. After all, the current in the medium is carried by drifting charge carriers around the quasi-Fermi level, which by scattering on the length scale of the carrier mean free path $\lambda_\mathrm{mfp}$ constantly adjust their momenta to the local current direction, and furthermore adjust their energies such that the local quasi-Fermi level follows the potential profile. We can therefore assume that also the energy and momentum relaxation at the interface between the 2D half-plane and the ballistic channel takes place on the length scale of $\lambda_\mathrm{mfp}$ in the 2D half-plane. On the drain side, an analogous situation occurs, with carriers exiting the ballistic channel with excess energy  that is quickly dissipated in the 2D half-plane (indicated by blue shading at $x'$ and $x^\ast$ in Fig.~\ref{fig:1}f of the Main Text) until carriers reach the local quasi-Fermi level that corresponds to the electrostatic potential (blue line). Also in this case, energy relaxation is accompanied by scattering to the appropriate momenta. Since the same scattering mechanisms are at work as on the source side, we can assume that carrier relaxation on the drain side also takes place in a thin strip close to the interface with a thickness on the order of $\lambda_\mathrm{mfp}$. Importantly, our approach assumes that the contacts are positioned appropriately such that Ohmic transport is ensured over the 2D half-plane and the distances $L$ and $W$ in our setup are much larger than the thickness of the non-Ohmic relaxation region (on the source and drain sides). In terms of length scales, we require that: $\lambda_\mathrm{mfp} \ll L$, $\lambda_\mathrm{mfp} \ll W$, $\lambda_\mathrm{mfp} \ll |\Phi_\mathrm{S,D}|/|\nabla \Phi|$.   

    Because carriers enter the ballistic channel at $-x'<0$ over a finite energy range $[+e\phi(-x'), -e\phi(-x')]$ (with a constant density of states in the channel), while arriving at the ballistic channel with an energy corresponding to the chemical potential $\mu(-x')$ in a thin strip of width $\lambda_\mathrm{mfp}$ next to the channel, and similarly leave the strip at $x'$ at the chemical potential $\mu(x')$ (see Fig.~\ref{fig:1}f of the Main Text), the chemical potential drops across this strip from $\mu(-x')=e\phi(-x')$ to $0$ on the source side and from $0$ to $\mu(+x')=-\mu(-x')=-e\phi(-x')$ on the drain side of the ballistic channel. This is fully analogous to the situation in the standard Landauer problem of local injection (Fig.~\ref{fig:1}e of the Main Text). Evidently, this creates a contact resistance in this strip, split equally between the source ($x<0$) and drain ($x>0$) sides of the ballistic channel. Connected to the associated voltage drop is a heat dissipation of $2e|\phi(x)|$, which is also distributed symmetrically between the source and drain sides. Importantly, as we have argued above, both contact resistance and heat dissipation occur within $\lambda_\mathrm{mfp}$ of the interface between the ballistic channel and the 2D half-plane. The situation is therefore similar to the classical Landauer picture of local injection, where contact resistance and heat dissipation also arise within  $\lambda_\mathrm{mfp}$ (in that case: of the contact material) at the point contacts. Thus, the 2D half-plane effectively acts as a continuation of the contact electrodes. However, because of it being extended, we must solve the Poisson equation to calculate the current density distribution which provides the injection current to the ballistic channel (and transports carriers away from it). In the Ohmic medium of the 2D half-plane this leads to a potential distribution that is linked to a continuous voltage drop in the 2D half-plane and a corresponding dissipation. As a result, on each path from the source to the drain electrode via the ballistic channel (entry point $-x'$ and exit point $x'$), there are two contributions to the overall energy dissipation, $2e|\phi(x')|$ occurring within $\lambda_\mathrm{mfp}$ of the interface to the channel, and $\mu_S-\mu_D - 2e|\phi(x')|$ spread over the entire path through the 2D half-plane, such that for all carriers the total dissipation irrespective of the path they take is always $\mu_{S}-\mu_{D}$, as it must. 

    Summarizing, we naturally arrive at the filling condition of Eq.~\eqref{eq:filling-condition} of the Main Text, which generalizes the standard Landauer problem to a setup with distributed injection, by imposing Ohmic transport on the half-plane and current continuity, further taking into account the large mismatch of mode number at the interface between the half-plane and the ballistic channel and ballistic propagation of charge carriers from source to drain regions. While ignoring the detailed scattering mechanisms that are involved in the transfer across the interface between the 2D half-plane and the ballistic channel (including non-Ohmic relaxation mechanisms), our argumentation has shown that these scattering mechanisms will be supported by the Ohmic medium of the 2D half-plane.      

	\section{Potential for classical quasi-1D Ohmic channel}
	\label{sec:appendixC}
	
	\subsection{General solution}
	
\begin{figure}[htb]
		\includegraphics[width=0.5\linewidth]{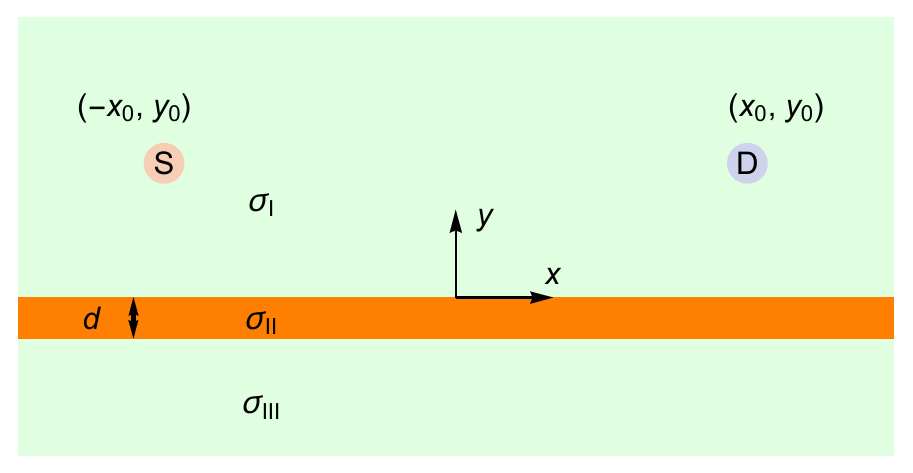}
		\caption{
			Configuration of a quasi-1D Ohmic channel (width $d \ll L,W$ and sheet conductivity $\sigma_\mathrm{II}$) sandwiched between two half-planes (with sheet conductivities $\sigma_\mathrm{I}$ and $\sigma_\mathrm{III}$ for the upper and lower half-planes, respectively). Two current injecting/extracting contacts are placed symmetrically on the upper half-plane. The resulting potential in the uppper half-plane is calculated.
		}
		\label{fig:S4}
	\end{figure}

	In this section the analytic solution of the problem of a classical quasi-1D Ohmic strip next to an Ohmic half-plane, sketched in Fig.~\ref{fig:S4}, is presented. We solve the corresponding boundary value problem. In fact, we consider a more general situation in which the half-plane below the quasi-1D Ohmic channel is taken explicitly into account, too. Its sheet conductivity is $\sigma_\mathrm{III}$. Apart from its broader applicability, this generalization has the advantage that no \textit{a priori} assumptions regarding the boundary condition at the lower edge of the quasi-1D channel are necessary.     
	
	The ansatz for the electrostatic potential in the three regions I, II, and III is 
	\begin{equation}
		\label{eq:Phi_I_ansatz}
		\Phi_\mathrm{I} (x, y) = \frac{I_{SD}}{4\pi \sigma_\mathrm{I}} \ln \left(\frac{(x - x_0)^2 + (y - y_0)^2}{(x + x_0)^2 + (y - y_0)^2}\right) + \frac{I_{SD}}{4\pi \sigma_\mathrm{I}} \int\limits_{0}^{\infty} A(k) e^{-k(y+d)} \sin(kx) dk,
	\end{equation}
	
	\begin{equation}
		\label{eq:Phi_II_ansatz}
		\Phi_\mathrm{II} (x, y) = \frac{I_{SD}}{4\pi \sigma_\mathrm{I}} \ln \left(\frac{(x - x_0)^2 + (y - y_0)^2}{(x + x_0)^2 + (y - y_0)^2}\right) + \frac{I_{SD}}{4\pi \sigma_\mathrm{I}} \int\limits_{0}^{\infty} \left[B(k) e^{k(y+d)} + C(k) e^{-k(y+d)}\right] \sin(kx) dk, 
	\end{equation}
	
	\begin{equation}
		\label{eq:Phi_III_ansatz}
		\Phi_\mathrm{III} (x, y) = \frac{I_{SD}}{4\pi \sigma_\mathrm{I}} \ln \left(\frac{(x - x_0)^2 + (y - y_0)^2}{(x + x_0)^2 + (y - y_0)^2}\right) + \frac{I_{SD}}{4\pi \sigma_\mathrm{I}} \int\limits_{0}^{\infty} D(k) e^{k(y+d)} \sin(kx) dk.
	\end{equation}
	The first term in Eq.~\eqref{eq:Phi_I_ansatz} follows from the Poisson equation for stationary currents in a 2D Ohmic sheet with sheet conductivity $\sigma_\mathrm{I}$ (in the Main Text, this sheet conductivity is labelled $\sigma$ for simplicity), when injecting a current $I_{SD}$ at point $(-x_0, y_0)$ and withdrawing it at $(+x_0, y_0)$. Its derivation can, e.g., be found in Ref.~\cite{Leis2022b}. Note that in the present case  $x_0=L/2$ and $y_0=W$. However, for better readability we will continue to use $x_0$ and $y_0$ in this section. Also note that the potentials in Eq.~\eqref{eq:Phi_I_ansatz}~to~\eqref{eq:Phi_III_ansatz} have opposite signs as the one in Ref.~\cite{Leis2022b}, because the current direction is reversed. The integral in Eq.~\eqref{eq:Phi_I_ansatz} embodies the influence of regions II and III on the potential distribution in region I. Its antisymmetry under the transformation $x\rightarrow -x$ follows from the symmetry properties of the problem. In regions II and III, the potential due to the current injection and extraction in region I is modified by boundary conditions at $y=0$ and $y=-d$ and by the different sheet conductivities in the two regions. This modification is captured by the integrals in Eq.~\eqref{eq:Phi_II_ansatz} and~\eqref{eq:Phi_III_ansatz}, which have the same antisymmetric property as Eq.~\eqref{eq:Phi_I_ansatz}, but differ in their dependence on $y$. Note that  $\lim_{y\rightarrow \infty} \Phi_\mathrm{I}(x, y)=0$ and  $\lim_{y\rightarrow -\infty} \Phi_\mathrm{III} (x, y)=0$, as it must. 
	
	The functions $A(k)$, $B(k)$, $C(k)$, and $D(k)$ follow from the four conditions of continuity regarding the potential $\Phi(x,y)$ and the current density $\sigma \partial\Phi(x,y)/\partial y$ at the boundaries at $y=0$ and $y=-d$. Specifically, we find
	\begin{equation}
		\label{eq:continuity_potential_d}
		A(k) e^{-kd} - B(k) e^{kd} - C(k) e^{-kd} = 0
	\end{equation}
	
	\begin{equation}
		\label{eq:continuity_potential_0}
		B(k) + C(k) - D(k) = 0
	\end{equation}
	
	\begin{equation}
		\label{eq:continuity_currentdensity_d}
		\int\limits_{0}^{\infty} k [\sigma_\mathrm{I} A(k) e^{-kd} + \sigma_\mathrm{II} B(k) e^{kd} - \sigma_\mathrm{II} C(k) e^{-kd}] \sin(kx) dk 
		= -2(\sigma_\mathrm{I} - \sigma_\mathrm{II}) \left[ \frac{y_0}{(x - x_0)^2 + y_0^2} - \frac{y_0}{(x + x_0)^2 + y_0^2} \right]
	\end{equation}
	
	\begin{equation}
		\label{eq:continuity_currentdensity_0}
		\int\limits_{0}^{\infty} k [\sigma_\mathrm{II} B(k) - \sigma_\mathrm{II} C(k) - \sigma_\mathrm{III} D(k)] \sin(kx) dk 
		= 2(\sigma_\mathrm{II} - \sigma_\mathrm{III}) \left[ \frac{y_0+d}{(x - x_0)^2 + (y_0+d)^2} - \frac{y_0+d}{(x + x_0)^2 + (y_0+d)^2} \right]
    \end{equation}
    Applying the Fourier sine transform 
    \begin{equation}
	   f(x) = \frac{2}{\pi} \int\limits_{0}^{\infty} \left[\int\limits_{0}^{\infty} f(x^\prime) \sin(x^\prime y) dx^\prime\right] \sin (xy) dy 
    \end{equation}
    appropriate for odd functions in the argument, to Eqs.~\eqref{eq:continuity_currentdensity_d} and \eqref{eq:continuity_currentdensity_0}, we obtain
    \begin{equation}
	   \label{eq:continuity_currentdensity_d_FT}
	   \sigma_\mathrm{I} A(k) e^{-kd} + \sigma_\mathrm{II} B(k) e^{kd} - \sigma_\mathrm{II} C(k) e^{-kd}
	   = - \frac{4(\sigma_\mathrm{I} - \sigma_\mathrm{II})}{\pi k} \int\limits_{0}^{\infty} \left[ \frac{y_0}{y_0^2 + (x - x_0)^2} - \frac{y_0}{y_0^2 + (x + x_0)^2} \right] \sin(kx) dx
    \end{equation}
    and
    \begin{equation}
	   \label{eq:continuity_currentdensity_0_FT}
	   \sigma_\mathrm{II} B(k) - \sigma_\mathrm{II} C(k) - \sigma_\mathrm{III} D(k) 
	   = \frac{4(\sigma_\mathrm{II} - \sigma_\mathrm{III})}{\pi k} \int\limits_{0}^{\infty} \left[ \frac{y_0+d}{(y_0+d)^2 + (x - x_0)^2} - \frac{y_0+d}{(y_0+d)^2 + (x + x_0)^2} \right] \sin(kx) dx,
	\end{equation}
    the right hand sides of which can be integrated (see p.\ 459, no.\ 3.732-1 in Ref.~\cite{gradshteyn2014table}) for $\beta > 0$ ($y_0 > 0$)
    \begin{equation}
	   \label{Gradstein_1}
	   \int\limits_{0}^{\infty} \left[ \frac{\beta}{\beta^2 + (x - y)^2} - \frac{\beta}{\beta^2 + (x + y)^2} \right] \sin( a x) dx = \pi  e^{- a \beta}  \sin(a \gamma)
	\end{equation}
    to yield
    \begin{equation}
	   \label{eq:continuity_currentdensity_d_FT_integrated}
	   \sigma_\mathrm{I} A(k) e^{-kd} + \sigma_\mathrm{II} B(k) e^{kd} - \sigma_\mathrm{II} C(k) e^{-kd} 
	   = - 4(\sigma_\mathrm{I} - \sigma_\mathrm{II}) e^{-k y_0} \frac{\sin(kx_0)}{k}
	\end{equation}
    and
    \begin{equation}
	   \label{eq:continuity_currentdensity_0_FT_integrated}
	   \sigma_\mathrm{II} B(k) - \sigma_\mathrm{II} C(k) - \sigma_\mathrm{III} D(k) = 4(\sigma_\mathrm{II} - \sigma_\mathrm{III}) e^{-k(y_0+d)} \frac{\sin(kx_0)}{k}. 
    \end{equation}
    Combining Eqs.~\eqref{eq:continuity_potential_0} and \eqref{eq:continuity_currentdensity_0_FT_integrated}, we get
    \begin{equation}
	   \label{eq:continuity_currentdensity_0_FT_integrated_prime}
	   (\sigma_\mathrm{II} - \sigma_\mathrm{III}) B(k) - (\sigma_\mathrm{II} + \sigma_\mathrm{III}) C(k) = 4(\sigma_\mathrm{II} - \sigma_\mathrm{III}) e^{-k(y_0+d)} \frac{\sin(kx_0)}{k}. 
    \end{equation}
    With Eqs.~\eqref{eq:continuity_potential_d}, \eqref{eq:continuity_currentdensity_d_FT_integrated}, and \eqref{eq:continuity_currentdensity_0_FT_integrated_prime}, we have derived a linear system of equations for the three functions $A(k)$, $B(k)$ and $C(k)$. Once these are known, $D(k)$ follows trivially from Eq.~\eqref{eq:continuity_potential_0}.

    Since we are mostly interested in region I and its boundary, which determines the injection potential and injection current density into the quasi-1D Ohmic channel (region II), we concentrate on $A(k)$ here. Solving the linear system of equations yields
    \begin{equation}
	   A(k) = -4 \frac{s_1 e^{2kd} + s_2}{1 + s_1s_2 e^{-2kd}} e^{-k(y_0+d)} \frac{\sin(kx_0)}{k},
    \end{equation}
    where we have defined 
    \begin{equation}
	   \label{eq:definition_s1_s2}
	   s_1 \equiv \frac{\sigma_\mathrm{I}-\sigma_\mathrm{II}}{\sigma_\mathrm{I}+\sigma_\mathrm{II}} \quad \quad \mathrm{and} \quad \quad s_2 \equiv \frac{\sigma_\mathrm{II}-\sigma_\mathrm{III}}{\sigma_\mathrm{II}+\sigma_\mathrm{III}}.
    \end{equation}
    Applying the series expansion $(1 + x)^{-1} = 1 - x+ x^2 - x^3 + ...$, this can be rewritten as 
    \begin{equation}
	   \begin{split}
		  A(k) &=-4 \sum_{n=0}^\infty(-s_1s_2)^n (s_1e^{-k2(n-1)d} + s_2e^{-k2nd}) e^{-k(y_0+d)} \frac{\sin(kx_0)}{k}\\
		  & = -4 \left[s_1 e^{-k(y_0 - 2d)} + (1 - s_1^2) s_2 \sum_{n=0}^\infty(-s_1s_2)^n e^{-k(y_0 + 2nd)}\right] e^{-kd}\frac{\sin(kx_0)}{k},
	   \end{split}
    \end{equation}
    where in the second line we have re-arranged the summands. Inserting this last equation into Eq.~\eqref{eq:Phi_I_ansatz} finally yields
    \begin{equation}
	   \label{eq:Phi_I_final}
	   \begin{split}
		  \Phi_\mathrm{I} (x, y)  & = \frac{I_{SD}}{4\pi \sigma_\mathrm{I}} \ln\left(\frac{(x - x_0)^2 + (y - y_0)^2}{(x + x_0)^2 + (y - y_0)^2}\right) + \frac{I_{SD}}{4\pi \sigma_\mathrm{I}} s_1 \ln\left(\frac{(x - x_0)^2 + (y + y_0)^2}{(x + x_0)^2 + (y + y_0)^2}\right) \\& + \frac{I_{SD}}{4\pi \sigma_\mathrm{I}}(1 - s_1^2)s_2 \sum^\infty_{n=0}(-s_1s_2)^n \ln\left(\frac{(x - x_0)^2 + [y + y_0 + 2(n+1)d]^2}{(x + x_0)^2 + [y + y_0 + 2(n+1)d]^2}\right),
	   \end{split}
    \end{equation}
    where we have used the integral
    \begin{equation}
	   \int\limits_{0}^{\infty} e^{-\beta x}\frac{\sin(\gamma x) \sin(ax)}{x} \ dx = \frac{1}{4} \ln \left( \frac{\beta^2 + (a + \gamma)^2}{\beta^2 + (a - \gamma)^2} \right)
    \end{equation}
    from Ref.~\cite{gradshteyn2014table} (see p.\ 537, no.\ 3.947). Eq.~\eqref{eq:Phi_I_final} has a simple physical interpretation: The first term is the contribution of the externally applied source and drain at $+y_0$ to the potential in region I, while the second term is the contribution to $\Phi_\mathrm{I}$ of an image source and drain due to mirroring at the interface between regions I and II  (hence the factor $s_1$); they are located at $-y_0$. Similarly, the $n=0$ term of the sum derives from a pair of images (source and drain) produced by the interface between the regions II and III (factor $s_2$), situated at $-y_0-2d$. The terms for $n>1$ are generated by mirroring the image source and drain deriving from the $n-1$ term twice, first at the far interface I-II, thereby picking up a factor $-s_1$, and then back at the adjacent interface II-III interface, collecting a factor $s_2$. This generates pairs of images at positions $-y_0-2(n+1)d$, farther and farther away from the two boundaries. The potential in region I thus originates from an infinite series of image sources and drains in regions II and III. Note that the factor $1-s_1^2$ in the third term of Eq.~\eqref{eq:Phi_I_final} takes care of the perfect screening that occurs at the interface I-II if the conductivity in region II is infinite. Then, no mirroring can take place at the far interface, because the adjacent one (to a perfect metal) completely screens it out; thus, the whole sum, including the $n=0$ term, must collapse, which is ensured by the above factor, as it becomes zero for $\sigma_\mathrm{II}\rightarrow\infty$. Eq.~\eqref{eq:Phi_I_final} was used to compute the cases labelled $\Omega_1$, $\Omega_2$ and $\Omega_3$ in the Main Text.      

    Based on Eq.~\eqref{eq:Phi_I_final}, the current density that is injected from region I into the quasi-1D Ohmic channel (region II) is given by
    \begin{equation}
	   \begin{split}
		  \frac{dI_\mathrm{inj}(x)}{dx}  = &\sigma_\mathrm{I} \left. \frac{\partial\Phi_\mathrm{I}(x,y)}{\partial y} \right|_{y=0^+} = \frac{I_{SD}}{4 \pi}  \left( - (1 - s_1) \frac{8 x_0 y_0 x}{[(x - x_0)^2 + y_0^2][(x + x_0)^2 + y_0^2]} \right. \\
		  & \phantom{(} \left. + (1 - s_1^2) s_2 \sum_{n=0}^{+\infty} (-s_1 s_2)^n \frac{8 x_0 [y_0 + 2 (n+1) d] x}{\{(x - x_0)^2 + [y_0 + 2 (n+1) d]^2\}\{(x + x_0)^2 + [y_0 + 2 (n+1) d]^2\}} \right).
	   \end{split}
    \end{equation}

    \subsection{Special cases}

    \subsubsection{Classical quasi-1D Ohmic edge channel}
    For an insulating region III ($\sigma_\mathrm{III}=0$, leading to $s_2 = 1$), Eq.~\eqref{eq:Phi_I_final} becomes 
    \begin{equation}
	   \label{eq:Phi_I_final_III_insulating}
	   \begin{split}
		  \Phi_\mathrm{I} (x, y)  & = \frac{I_{SD}}{4\pi \sigma_\mathrm{I}} \ln\left(\frac{(x - x_0)^2 + (y - y_0)^2}{(x + x_0)^2 + (y - y_0)^2}\right) + \frac{I_{SD}}{4\pi \sigma_\mathrm{I}} s_1 \ln\left(\frac{(x - x_0)^2 + (y + y_0)^2}{(x + x_0)^2 + (y + y_0)^2}\right) \\& + \frac{I_{SD}}{4\pi \sigma_\mathrm{I}}(1 - s_1^2) \sum^\infty_{n=0}(-s_1)^n \ln\left(\frac{(x - x_0)^2 + (y + y_0 + 2(n+1)d)^2}{(x + x_0)^2 + (y + y_0 + 2(n+1)d)^2}\right).
	   \end{split}
    \end{equation}
    This equation describes the situation of a classical quasi-1D edge channel, i.e., a channel with sheet conductivity $\sigma_\mathrm{II}$ that is connected on one of its sides to an Ohmic half-plane with sheet conductivity $\sigma_\mathrm{I}$, on the other side to an insulator. The corresponding current density injected into the quasi-1D Ohmic channel then reads
    \begin{equation}
	   \label{eq:injection_current_I_final_III_insulating}
	   \begin{split}
		  \frac{dI_\mathrm{inj}(x)}{dx} &= \sigma_\mathrm{I} \left. \frac{\partial\Phi_\mathrm{I}(x,y)}{\partial y} \right|_{y=0^+} = \frac{I_{SD}}{4 \pi}  \left( - (1 - s_1) \frac{8 x_0 y_0 x}{[(x - x_0)^2 + y_0^2][(x + x_0)^2 + y_0^2]} \right. \\
		  & \phantom{(} \left. + (1 - s_1^2) \sum_{n=0}^{\infty} (-s_1)^n \frac{8 x_0 [y_0 + 2 (n+1) d] x}{\{(x - x_0)^2 + [y_0 + 2 (n+1) d]^2\}\{(x + x_0)^2 + [y_0 + 2 (n+1) d]^2\}} \right).
	   \end{split}
    \end{equation}
    Eqs.~\eqref{eq:Phi_I_final_III_insulating} and \eqref{eq:injection_current_I_final_III_insulating} were used to generate the results for the quasi-1D Ohmic channels $\Omega_1$, $\Omega_2$, and $\Omega_3$.

    \subsubsection{Classical quasi-1D Ohmic channel with infinite sheet conductivity} \label{subsubsec:pc}
    We now consider several special cases of Eq.~\eqref{eq:Phi_I_final} that are used in the Main Text. First, we consider an `Ohmic' quasi-1D channel with a sheet conductivity $\sigma_\mathrm{II}\rightarrow \infty $ and injection through a 2D half-plane with finite conductivity $\sigma_\mathrm{I}$. In a sense, this can be understood as the resistive analog of a ballistic channel with quantized conductance and distributed injection as treated in the Main Text. In this limit, $s_1\rightarrow -1$ and $s_2\rightarrow 1$, and we obtain
    \begin{equation}
	   \label{eq:Phi_I_final_sigma2_infinite}
	   \begin{split}
		  \Phi_\mathrm{I} (x, y)  & = \frac{I_{SD}}{4\pi \sigma_\mathrm{I}} \ln\left(\frac{(x - x_0)^2 + (y - y_0)^2}{(x + x_0)^2 + (y - y_0)^2}  \cdot  \frac{(x + x_0)^2 + (y + y_0)^2}{(x - x_0)^2 + (y + y_0)^2}\right),
	   \end{split}
    \end{equation}
    This equation was used to compute the case $\sigma_\Omega \rightarrow \infty$ in the Main Text. Notably, there is no dependence on the width of the perfectly conducting channel. Also, we see that the infinite conductivity of the quasi-1D channel effectively screens out the influence of region III on the potential distribution in region I (no influence of $\sigma_\mathrm{III}$ on $\Phi_\mathrm{I}$). This also holds if image source and drain at $-y_0$ are located within region III. 

    In the $\sigma_\mathrm{II} \rightarrow \infty$ limit, we obtain $\Phi_\mathrm{I}(x,y=0^+)=0$ and the current density of injection is given by
    \begin{equation}
	   \label{eq:injection_current_pc}
	   \sigma_\mathrm{I} \partial_y \Phi_\mathrm{I}(x,y)|_{y=0^+} = -\frac{I_{SD}}{4 \pi } \frac{16 x_0 y_0 x}{[(x - x_0)^2 + y_0^2][(x + x_0)^2 + y_0^2]},
    \end{equation}
    which corresponds to the first term on the left-hand side of our filling condition (Eq.~\eqref{eq:filling-condition-alt}). At the position $-x^\ast = -\sqrt{x_0^2 + y_0^2}$ (see below), the injection current density is equal to
    \begin{equation}
	   \sigma_\mathrm{I} \partial_y \Phi_\mathrm{I}(-x^\ast,y)|_{y=0^+} = \frac{I_{SD}}{4 \pi} \frac{4 x_0}{y_0 x^\ast} 
    \end{equation}
    which is used to normalize the current density in Fig.~\ref{fig:S2}b. Note that this does not correspond to the maximal current density, as can be seen in Fig.~\ref{fig:S2}b, which is obtained at $x = -(x_0^2 - y_0^2 + 2 \sqrt{x_0^4 + x_0^2 y_0^2 + y_0^4})^{1/2}/\sqrt{3}$.

    \subsubsection{Classical quasi-1D Ohmic channel with vanishing conductivity}
    In the limit $\sigma_\mathrm{II} \rightarrow 0$ (referred to as the $\sigma_\Omega \rightarrow 0$ limit in the Main Text), Eq.~\eqref{eq:Phi_I_final} of the classical quasi-1D channel becomes
    \begin{equation}
	   \label{eq:Phi_I_final_sigma2_zero}
		  \Phi_\mathrm{I} (x, y)  = \frac{I_{SD}}{4\pi \sigma_\mathrm{I}} \ln\left(\frac{(x - x_0)^2 + (y - y_0)^2}{(x + x_0)^2 + (y - y_0)^2}  \cdot  \frac{(x - x_0)^2 + (y + y_0)^2}{(x + x_0)^2 + (y + y_0)^2}\right).
    \end{equation}
    At the interface, i.e., for $y=0^+$, 
    \begin{equation}
	   \label{eq:Phi_I_final_sigma2_zero_y_zero}
	   \Phi_\mathrm{I} (x, y=0^+)  = \frac{I_{SD}}{4\pi \sigma_\mathrm{I}} 2\ln\left(\frac{(x - x_0)^2 + y_0^2}{(x + x_0)^2 + y_0^2} \right).
    \end{equation}
    The two extrema of this function are found at $\pm x^\ast \equiv \pm \sqrt{x_0^2 + y_0^2}$, the corresponding extremal values are 
    \begin{equation} \label{eq:phi-ast_vanishing-conductivity}
	   \pm\phi^\ast = \pm \frac{I_{SD}}{4 \pi \sigma_I } 2 \ln\left( \frac{x^\ast + x_0}{x^\ast - x_0} \right), 
    \end{equation}
    where the upper signs relate to the maximum at $-x^\ast$, which we used to normalize the interface potentials in Fig.~\ref{fig:2} of the Main Text and Figs.~\ref{fig:S1}-\ref{fig:S3}. As expected, from Eq.~\eqref{eq:Phi_I_final_sigma2_zero} we obtain $\sigma_\mathrm{I}\partial_y \Phi_\mathrm{I}(x,y)|_{y=0^+} = 0$ for the injection current density into the insulating quasi-1D channel.   

    \subsubsection{Classical Ohmic lower half-plane}
    In the Main Text and \ref{sec:appendixB} we also compare the ballistic edge channel to an Ohmic half-plane with finite sheet conductivity as a proxy (case labeled `lhp'). Starting from equation Eq.~\eqref{eq:Phi_I_final}, the latter can naturally be reached by taking the limit $d\rightarrow \infty $, whence we get
    \begin{equation}
	   \label{eq:Phi_I_final_d=infty}
        \Phi_\mathrm{I} (x, y)  = \frac{I_{SD}}{4\pi \sigma_\mathrm{I}} \ln\left(\frac{(x - x_0)^2 + (y - y_0)^2}{(x + x_0)^2 + (y - y_0)^2} \right) + \frac{\sigma_\mathrm{I}-\sigma_\mathrm{II}}{\sigma_\mathrm{I}+\sigma_\mathrm{II}} \frac{I_{SD}}{4\pi \sigma_\mathrm{I}} \ln\left(\frac{(x - x_0)^2 + (y + y_0)^2}{(x + x_0)^2 + (y + y_0)^2} \right).
    \end{equation}
    We note that this is the same expression as Eq.~(8) in Ref.~\cite{Leis2022b}. Evidently, it can also be obtained by taking the limit $d\rightarrow 0$ and assuming, without loss of generality, $\sigma_\mathrm{II}=\sigma_\mathrm{III}$ ($s_2=0$), or indeed assuming $\sigma_\mathrm{II}=\sigma_\mathrm{I}$ ($s_1=0$), noting in the latter case that $(-s_1s_2)^0$ remains finite and equal to 1 even if $s_1\rightarrow 0$. The extrema $\pm\phi^\ast$ of Eq.~\eqref{eq:Phi_I_final_d=infty} can again be worked out analytically and are located at $\pm x^\ast$, with  
    \begin{equation}
	   \phi^\ast = \frac{I_{SD}}{4 \pi } \frac{2}{\sigma_\mathrm{I} + \sigma_\mathrm{II}}  \ln\left( \frac{x^\ast + x_0}{x^\ast - x_0} \right). 
    \end{equation}
    The current density of injection is given by
    \begin{equation}
	   \sigma_\mathrm{I} \partial_y \Phi_\mathrm{I}(x,y)|_{y=0^+} = - \frac{I_{SD}}{4 \pi} \frac{\sigma_\mathrm{II}}{\sigma_\mathrm{I} + \sigma_\mathrm{II}} \frac{16 x_0 y_0 x}{[(x-x_0)^2 + y_0^2][(x+x_0)^2 + y_0^2]}.
    \end{equation}
    If we additionally assume perfect conductivity ($\sigma_\mathrm{II}\rightarrow \infty$) for the `lhp' case in Eq.~\eqref{eq:Phi_I_final_d=infty}, we find 
    \begin{equation}
	   \label{eq:Phi_I_final_d=0_sigma3_infinite}
		\Phi_\mathrm{I} (x, y)  = \frac{I_{SD}}{4\pi \sigma_\mathrm{I}} \ln\left(\frac{(x - x_0)^2 + (y - y_0)^2}{(x + x_0)^2 + (y - y_0)^2}  \cdot  \frac{(x + x_0)^2 + (y + y_0)^2}{(x - x_0)^2 + (y + y_0 )^2}\right).
    \end{equation}
    The solution Eq.~\eqref{eq:Phi_I_final_d=0_sigma3_infinite} for a lower half-plane with infinite sheet conductivity is in fact identical to the one for the classical quasi-1D channel with infinite conductivity ($\sigma_\Omega \rightarrow \infty$) in Eq.~\eqref{eq:Phi_I_final_sigma2_infinite}.

    \section{Experimental estimations}
	\label{sec:appendixD}

    We provide a quantitative estimate for a realistic STM-based multi-tip potentiometry experiment. The estimation demonstrates that this technique can detect signatures of a ballistic channel in the interface potential $\phi(x)$, and in particular in its maximum value $\phi^\ast$ (see, e.g., Fig.~\ref{fig:1}d in the Main Text). Assuming the point-like source and drain contacts in Fig.~1b to have a physical diameter $D$ (with $D \ll W, L$) and a potential $\pm V_\text{c}$ at their circumference, 
    we obtain for $V_\mathrm{c}$
    \begin{equation}
     V_\text{c} \approx \frac{I_{SD}}{4 \pi \sigma} \ln\left( \frac{4 L^2}{D^2} \cdot \frac{4 W^2}{L^2 + 4 W^2} \right).
    \end{equation}
    To derive this equation, we inserted $(x, y) = (-L/2+D/2, W)$ on the source side and $(x, y) = (L/2-D/2, W)$ on the drain side into the expression for $ \Phi_{SD}(x,y)$, given in Eq.~\eqref{eq:Phi_SD}.  We note that Eq.~\eqref{eq:Phi_SD} neglects the correction due to the interface potential $\phi_\mathrm{ball}$ (see Eq.~\eqref{Eq:sum}), but close to the contacts the latter's contribution is small (for not too small $W$).
    
    Next, we estimate the maximum $\phi^\ast$ of the interface potential as
    \begin{align}
         \label{eq:phi-ast-scaling}
        \begin{split}
            \phi^\ast &= \frac{\phi^\ast}{\phi^\ast_{\sigma_\Omega \rightarrow 0}} \times \frac{I_{SD}}{4 \pi \sigma} 2 \ln\left( \frac{\sqrt{L^2/4 + W^2} + L/2}{\sqrt{L^2/4 + W^2} - L/2} \right) \\
            &\approx \frac{\phi^\ast}{\phi^\ast_{\sigma_\Omega \rightarrow 0}} \times V_\text{c} \, 2 \left. \ln\left( \frac{\sqrt{L^2/4 + W^2} + L/2}{\sqrt{L^2/4 + W^2} - L/2} \right) \right/ \ln\left( \frac{4 L^2}{D^2} \cdot \frac{4 W^2}{L^2 + 4 W^2} \right).
        \end{split}
    \end{align}
    Here, we expressed the maximum $\phi^\ast$ of the interface potential in terms of the ratio  $\phi^\ast/\phi^\ast_{\sigma_\Omega\rightarrow 0}$, the applied potential $V_\mathrm{c}$, and the geometry parameters $D$, $L$, and $W$.  $\phi^\ast_{\sigma_\Omega\rightarrow 0}$ is the maximum of the interface potential for the Ohmic edge channel with vanishing conductivity, for which we have derived an analytic expression (Eq.~\eqref{eq:phi-ast_vanishing-conductivity}) and which provides a natural upper limit for  $\phi^\ast$, because it is the potential of a fully insulting edge of the upper half-plane. Any injection into the 1D channel will reduce this potential:   $\phi^\ast/\phi^\ast_{\sigma_\Omega \rightarrow 0} \leq 1$.
    
    Inserting typical values such as $D = 10\,\text{nm}$, $L = 1\,\text{\textmu m}$, $W=0.2\,\text{\textmu m}$, and $\sigma/(G_0/2) = 1/4$ ($\sigma \approx 10 \, \text{\textmu S}$), with $\phi^\ast/\phi^\ast_{\sigma_\Omega \rightarrow 0} \approx 0.15$ (see Fig.~\ref{fig:S3}d, for example), we obtain an interface potential maximum $\phi^\ast$ of $1.5$~to~$15 \,\text{mV}$ when applying a potential $V_\text{c} = 10$~to~$100\,\text{mV}$, respectively, to the contacts (corresponding to a source-to-drain current of $I_{SD} \approx 0.2$~to~$1.7 \, \text{\textmu A}$). Hence, the interface potential should be well within the resolution of STM-based potentiometry ($10$~to~$100\,\text{\textmu V}$)~\cite{Luepke2015}.
    
    From Eq.~\eqref{eq:phi-ast-scaling} and our numerical results for $\phi^\ast/\phi^\ast_{\sigma_\Omega \rightarrow 0}$ (Fig.~\ref{fig:2} of the Main Text and Figs.~\ref{fig:S3}a,d) we can see that $\phi^\ast$ is robust against changes of $L$, but will naturally become more difficult to resolve for larger $W$ (whence $\phi^\ast_{\sigma_\Omega \rightarrow 0} \rightarrow 0$ and hence also $\phi^\ast\rightarrow 0$, since  $\phi^\ast\leq \phi^\ast_{\sigma_\Omega \rightarrow 0}$) and smaller ratios of $\sigma/(G_0/2)$ (for which $\phi^\ast / \phi^\ast_{\sigma_\Omega \rightarrow 0} \rightarrow 0$, as Fig.~\ref{fig:2}b of the Main Text shows).

    \putbib[bu2.bbl]
    \end{bibunit}
    
\end{document}